\documentclass[a4paper,fleqn,usenatbib]{mnras}

\usepackage[T1]{fontenc}
\usepackage{times}

\usepackage{graphicx}	
\usepackage{amsmath}	
\usepackage{amssymb}	

\graphicspath{{./}{Figures/}}

\newcommand{\md}{\mathrm{d}}

\newcommand{\asd}{\bar\Sigma}

\def\ba{\begin{eqnarray}}
\def\ea{\end{eqnarray}}

\title[Radial Profiles of Debris discs]{Radial Profiles of Surface Density in Debris Discs}

\author[R. R. Rafikov]{
Roman R. Rafikov$^{1,2}$\thanks{E-mail: rrr@damtp.cam.ac.uk}\\
$^1$Department of Applied Mathematics and Theoretical Physics, University of Cambridge, Wilberforce Road, Cambridge CB3 0WA, UK\\
$^2$Institute for Advanced Study, Einstein Drive, Princeton, NJ 08540, USA}

\date{Accepted XXX. Received YYY; in original form ZZZ}

\pubyear{2020}

\begin{document}
\label{firstpage}
\pagerange{\pageref{firstpage}--\pageref{lastpage}}
\maketitle

\begin{abstract}
Resolved observations of debris discs can be used to derive radial profiles of Azimuthally-averaged Surface Density (ASD), which carries important information about the disc structure even in presence of non-axisymmetric features and has improved signal-to-noise characteristics. We develop a (semi-)analytical formalism allowing one to relate ASD to the underlying semi-major axis and eccentricity distributions of the debris particles in a straightforward manner. This approach does not involve the distribution of particle apsidal angles, thus simplifying calculations. It is a much faster, more flexible and effective way of calculating ASD than the Monte Carlo sampling of orbital parameters of debris particles. We present explicit analytical results based on this technique for a number of particle eccentricity distributions, including two cases of particular practical importance: a prescribed radial profile of eccentricity, and the Rayleigh distribution of eccentricities. We then show how our framework can be applied to observations of debris discs and rings for retrieving either the semi-major axis distribution or (in some cases) the eccentricity distribution of debris, thus providing direct information about the architecture and dynamical processes operating in debris discs. Our approach  also provides a fast and efficient way of forward modeling observations. Applications of this technique to other astrophysical systems, e.g. the nuclear stellar disc in M31 or tenuous planetary rings, are also discussed. 
\end{abstract}




\section{Introduction}
\label{sec:Introduction}

High-resolution observations of debris discs allow us to determine the spatial distribution of brightness contributed by debris particles in projection onto the sky plane. Brightness distribution can be converted to a spatial distribution of the surface density of emitting particles $\Sigma(r,\phi)$ in polar $(r,\phi)$ coordinates in the plane of the disc, using straightforward assumptions, e.g. the knowledge of particle emissivity as a function of $r$ (and $\phi$ in the case of a disc directly imaged in the near-IR or optical bands), information on disc inclination (and the possibility of it being warped), and so on \citep{Wyatt2008}. Such a conversion provides us with very important information about the instantaneous spatial  distribution of debris, which can be linked to processes that shaped the underlying planetary system in the first place. 

Inference of $\Sigma(r,\phi)$ from observations can also inform us about the eccentricities and inclinations of the emitting particles (as non-circular motion of the debris particles is another factor affecting $\Sigma(r,\phi)$), shedding light on the dynamical processes shaping the disc. This may allow one to distinguish dynamical shaping of the disc by the embedded planets via secular forcing \citep{Lee2016,Pearce2014,Sefilian2020} or through self-stirring \citep{KenW2010,Marino2021}.

The goal of this work is to present a mathematical formalism allowing one to easily connect the semi-major axis distribution of debris particles and their dynamical state (i.e. their eccentricity distribution) to a certain characteristic of $\Sigma(r,\phi)$ that can be inferred from observation. More specifically, focusing on the case of a flat, co-planar disc we define {\it Axisymmetric (or Azimuthally-averaged) Surface Density} $\asd$ (hereafter ASD) as the azimuthal average (zeroth-order azimuthal moment) of its $\Sigma(r,\phi)$:
\ba  
\asd(r)=\frac{1}{2 \pi}\int_0^{2\pi}\Sigma(r,\phi)\md\phi.
\label{eq:asd}
\ea  
Clearly, $\asd(r)$ can be easily determined once $\Sigma(r,\phi)$ is inferred from observations as described earlier. ASD has a higher signal-to-noise ratio than $\Sigma(r,\phi)$, making $\asd$ a metric of choice for detailed analysis in many studies of debris discs \citep{Marino2021}. 

In this work we provide a way to link $\asd(r)$ to the semi-major axis distribution of Keplerian particles comprising the disc, given the knowledge of their kinematic properties. We also explore the possibility of using the observationally inferred ASD to solve the inverse problem of determining the radial mass distribution in the disc, or the dynamical characteristics (i.e. the eccentricity distribution) of the debris particles. 

This work is organised as follows. After outlining our basic setup (\S\ref{sec:Setup}), we proceed to develop our formalism for discs with a prescribed profile of eccentricity (\S\ref{sec:unique_e}), with the master equation (\ref{eq:asd-rel-gen}) being the key result. We then extend our approach to the case of a continuous distribution of particle eccentricities (\S\ref{sec:e_dist}), quoting some explicit analytical results for the case of the Rayleigh distribution in \S\ref{sec:Rayleigh}; a number of other eccentricity distributions are explored in Appendices \ref{sec:psi-rings} \& \ref{sec:psi-edge}. We explore the use of our results for retrieving the semi-major axis distribution (\S\ref{sec:inverse-mass}) or the eccentricity distribution (\S\ref{sec:inverse-e}) of debris particles from observations. We discuss our findings in \S\ref{sec:disc} and summarise in \S\ref{sec:Summary}. A number of additional technical but useful results for certain types of debris discs and rings can be found in Appendices. 


\section{Basic Setup}
\label{sec:Setup}


We consider a disc of particles orbiting a central mass $M_\star$. We restrict ourselves to the planar geometry and employ polar $(r,\phi)$ coordinates. The orbits of objects comprising the disc are characterized by their semi-major axis $a$, eccentricity $e$ and apsidal angle $\varpi$ (relative to a fixed reference direction), and are given by  
\ba  
r=\frac{a(1-e^2)}{1+e\cos(\phi-\varpi)}.
\label{eq:ellipse}
\ea  
Keplerian motion of a particle with a non-zero eccentricity $e$ allows it to contribute to surface density at radii different from its semi-major axis, namely at $r$ satisfying  
\ba
r_p\le r\le r_a,
\label{eq:cond}
\ea
where $r_p=a(1-e)$ and $r_a=a(1+e)$ are the periastron and apoastron distances. Spatial overlap of a large number of disc particles moving along their orbits with different $a$, $e$ and $\varpi$ and random mean anomalies --- a natural assumption adopted in this work (as well as many others) --- determines the spatial distribution of the surface density $\Sigma(r,\phi)$, which is potentially observable.   

Let $\mu(a)$ be the semi-major axis distribution of mass in the disc, such that $\md m(a)=\mu(a)\md a$ is the amount of mass in disc particles having their semi-major axes in the interval $(a,a+\md a)$ \citep{Statler01}. We find it useful to define $\Sigma_a(a)$ such that 
\ba  
\Sigma_a(a)=\frac{1}{2\pi a}\frac{\md m}{\md a}=\frac{\mu(a)}{2\pi a}.
\label{eq:sigma_a}
\ea  
Note that $\Sigma_a(a)$ does not have a clear physical interpretation, but for a disc composed of objects on circular orbits one naturally has $a=r$ and $\Sigma_a(a)=\Sigma(r)$, with no $\phi$-dependence. 

Knowledge of $\Sigma_a(a)$ or $\mu(a)$ is important, since in some cases it might be linked to the initial distribution of mass in the disc. For example, if the disc evolves secularly under the action of massive external perturber(s), then the semi-major axes of its constituent particles would be conserved \citep{Murray1999}. As a result, both $\mu(a)$ and $\Sigma_a$ remain invariant in the course of secular evolution, and their knowledge at present time can inform us about the mass distribution in a debris disc at its formation.

We will first consider in \S\ref{sec:unique_e} a simple disc model in which $e$ is a {\it unique} function of $a$, i.e. a disc in which all particles with a semi-major axis $a$ have a single value of eccentricity $e_a(a)$. However, in a debris disc, particles having the same semi-major axis $a$ may also have different values of eccentricity and orbit crossings are allowed (unlike fluid discs for which $e$ is always a unique function of $a$). As a result, kinematic properties of a debris disc are best characterized by a {\it distribution} of eccentricities $\psi_e(e,a)$ such that $\psi_e(e,a)\md e$ is the fraction of particles having $e$ in the interval $(e,e+\md e)$ at a given $a$. We will explore this more general model in \S\ref{sec:e_dist}.

\begin{figure}
	\includegraphics[width=\linewidth]{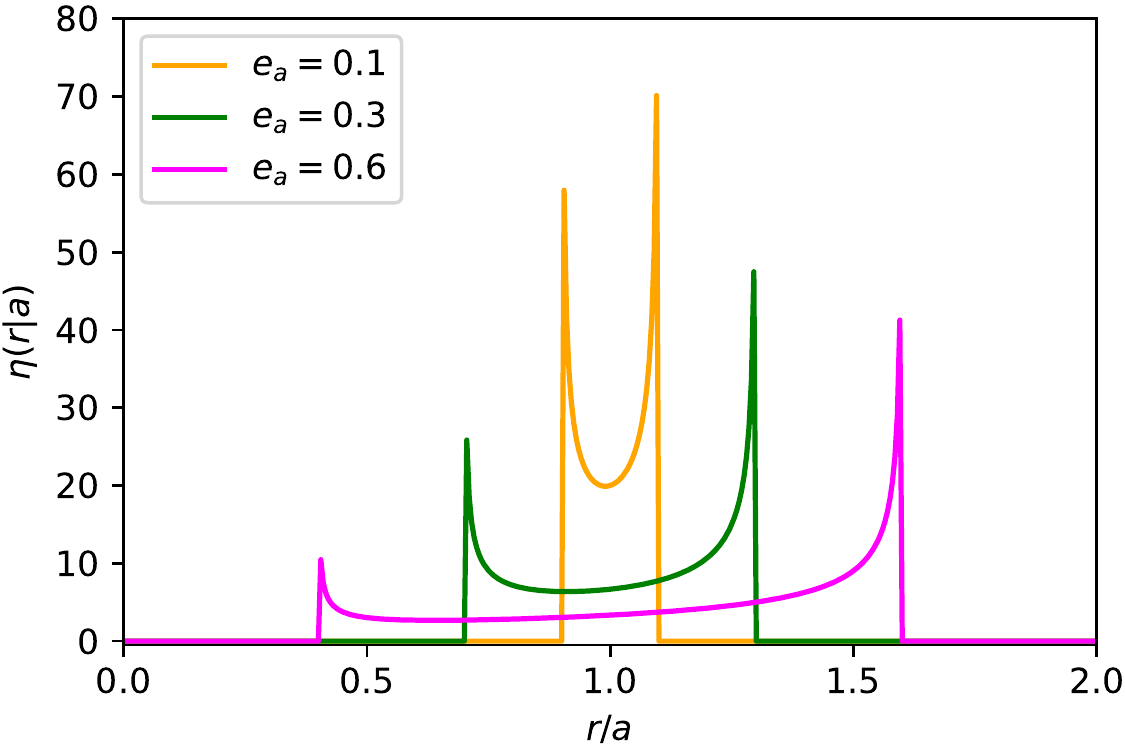}
    \caption{Behavior of $\eta(r|a)$ defined by the equation (\ref{eq:eta1}) as a function of $r/a$ for $e_a=0.1, 0.3, 0.6$; we assumed $\Sigma(a)=1$.}
    \label{fig:eta}
\end{figure}


\section{discs with $e=e(a)$}
\label{sec:unique_e}


Let us assume that eccentricity of the debris disc particles is a unique function of their semi-major axis, $e=e_a(a)$. This is equivalent to a disc having eccentricity distribution in the form
\ba 
\psi_e(e,a)=\delta(e-e_a(a)),
\label{eq:Dirac}
\ea 
where $\delta$ is the Dirac $\delta$-function.

We introduce an auxiliary function $\eta(r|a)$ such that $\eta(r|a)\md r\md a$ is the amount of mass per radial interval $(r,r+\md r)$ contributed by objects with eccentricity $e$ and semi-major axes in the interval $(a,a+\md a)$. It is clear that  $\eta(r|a)$ is non-zero only as long as the condition (\ref{eq:cond}) is fulfilled, with $r_p=r_p(a)$ and $r_a=r_a(a)$, where 
\ba 
r_p(a)=a(1-e_a(a)),~~~r_a(a)=a(1+e_a(a)).
\label{eq:peri-apo}
\ea  
Moreover, from the definition of $\eta(r|a)$ and mass conservation it immediately follows that the following normalization condition must be obeyed, see equation (\ref{eq:sigma_a}):
\ba
\md a\int_{r_p(a)}^{r_a(a)}\eta(r|a)\md r=\md m(a)=2\pi a\Sigma_a(a)\md a.
\label{eq:norm}
\ea  

It is clear on physical grounds that the amount of mass  $\eta(r|a)\md r$ that an orbiting particle contributes to the radial interval $(r,r+\md r)$, in the orbit-averaged sense, should be proportional to the amount of time $\md t$ that it spends in this radial range, i.e. $\eta(r|a)\md r\propto \md t$. Thus, $\eta(r|a)\propto \md t/\md r\propto v_r^{-1}$, where \citep{Murray1999}
\ba  
v_r=\sqrt{\frac{GM_\star}{a}}\frac{\sqrt{(r-r_p)(r_a-r)}}{r}
\label{eq:v_r}
\ea 
is the radial velocity of a particle with eccentricity $e$ and semi-major axis $a$ in a Keplerian potential. This implies that
\ba  
\eta(r|a)=\frac{C(a)r}{\sqrt{(r-r_p)(r_a-r)}},
\label{eq:eta}
\ea  
where $C(a)$ is a normalization constant, which still depends on $a$.

To fix $C(a)$ we plug the expression (\ref{eq:eta}) into the condition (\ref{eq:norm}), finding as a result $C(a)=2\pi I^{-1}a\Sigma_a(a)$, where
\ba  
I=\int_{r_p}^{r_a}\frac{r\md r}{\sqrt{(r-r_p)(r_a-r)}}=\pi a
\label{eq:I}
\ea  
is independent of $e$, because of the definitions of $r_p$ and $r_a$. Thus,
\ba  
\eta(r|a)=\frac{2r\Sigma_a(a)}{\sqrt{(r-r_p)(r_a-r)}}.
\label{eq:eta1}
\ea  
The behavior of $\eta(r|a)$ as a function of $r$ (for a fixed $a$) is illustrated in Figure \ref{fig:eta} for several values of $e_a$. 

Let us call $\eta(r)$ the amount of mass in debris particles per unit physical radius $\md r$ (and not $\md a$, distinguishing $\eta(r)$ from $\mu(a)$). Since the mass at a given radius $r$ is contributed by all particles with semi-major axes satisfying the condition (\ref{eq:cond}), $\eta(r)$ can be obtained by summing over all contributions $\eta(r|a)$ satisfying this condition. In other words, 
\ba  
\eta(r)=\int_0^\infty\eta(r|a)\md a = 2r \int_0^\infty\frac{\Sigma_a(a)\theta(r,a)\,\md a}{\sqrt{(r-r_p(a))(r_a(a)-r)}}.
\label{eq:etar}
\ea  
with 
\ba  
\theta(r,a)=\Theta((r-r_p(a))(r_a(a)-r)),
\label{eq:theta}
\ea  
where $\Theta(z)$ is a Heaviside step-function; $\theta=1$ for all $a$ such that $r_p(a)\le r\le r_a(a)$, while $\theta=0$ if this condition is violated. Introduction of the indicator function $\theta(r,a)$ forces us to count in the integral (\ref{eq:etar}) the contribution to the local surface density only from particles with orbits crossing the radius $r$. It also  constrains the expression inside the radical to be non-negative.

\begin{figure}
	\includegraphics[width=\linewidth]{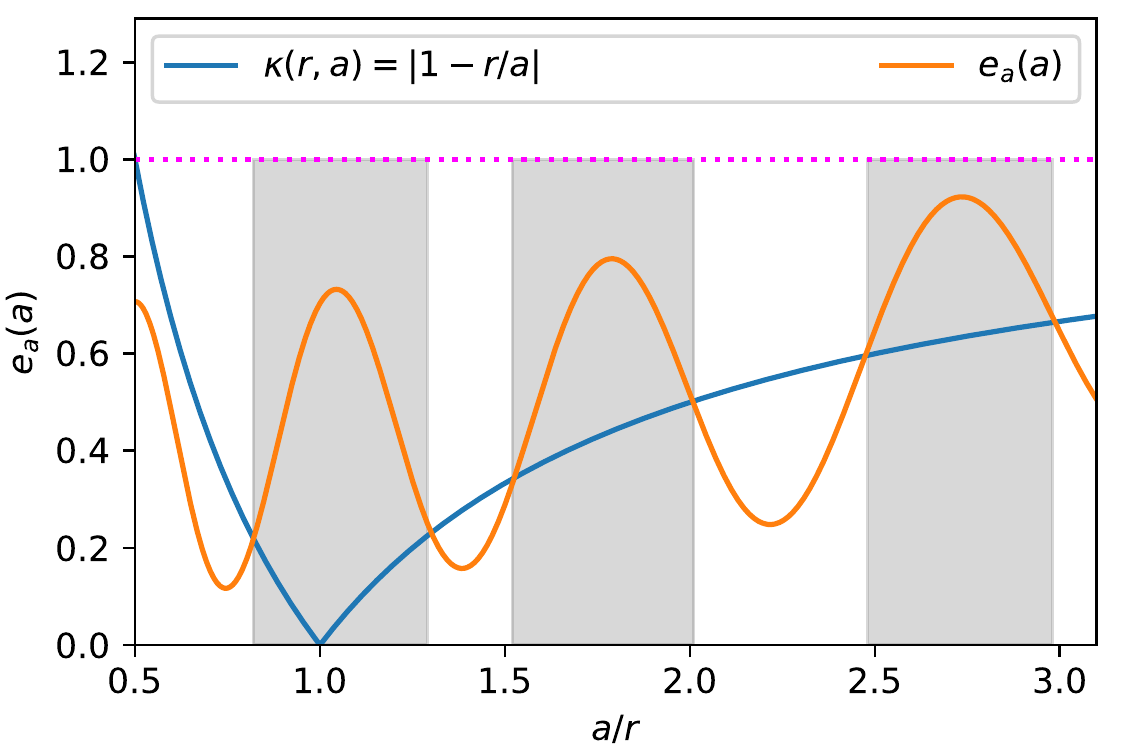}
    \caption{Schematic illustration showing the intervals of semi-major axis $a$ (grey shaded) which contribute to particle density at radius $r$. Blue curve shows $\kappa(r,a)$, see equation (\ref{eq:kappa}), while orange curve shows some $e_a(a)$ profile. Only the particles with $a$ such that $e_a(a)>\kappa(r,a)$ can reach $r$.}
    \label{fig:e-scheme}
\end{figure}

On the other hand, by its definition, $\eta(r)$ must coincide with $r\int_0^{2\pi}\Sigma(r,\phi)\md\phi=2\pi r\asd(r)$, see equation (\ref{eq:asd}). Expressing $\asd(r)=\eta(r)/(2\pi r)$, we obtain
\ba  
\asd(r)=\pi^{-1}\int_0^\infty\frac{\Sigma_a(a)\theta(r,a)}{\sqrt{(r-r_p(a))(r_a(a)-r)}} \md a,
\label{eq:asd-rel}
\ea  
which is the desired relation between $\asd(r)$ and $\Sigma_a$(a). One can easily show that $\asd(r)\to \Sigma_a(r)$ for $e_a(a)\to 0$, as expected.


\subsection{Master equation}
\label{sec:master-eq}


One can rewrite the result (\ref{eq:asd-rel}) in a different form, making more explicit the role of the constraint (\ref{eq:cond}). This constraint can be written as a {\it lower limit} on the eccentricity $e_a(a)$ of a particle with the semi-major axis $a$ capable of reaching radius $r$: $e_a(a)>1-(r/a)$ for $r<a$, and $e_a(a)>(r/a)-1$ for $r>a$, see Figure \ref{fig:e-scheme} for an illustration. Alternatively, this can be written as
\ba  
e_a(a)>\kappa(r,a), ~~~\mbox{where}~~~ \kappa(r,a)=\left|1-\frac{r}{a}\right|,
\label{eq:kappa}
\ea  
a constraint valid for both $r<a$ and $r>a$. Since $e_a(a)<1$, equation (\ref{eq:kappa}) necessarily implies that $a>r/2$. Indeed, particles with $a<r/2$ have their apoastra $r_a(a)<r$ (i.e. do not reach $r$) even for $e_a(a)\to 1$. For that reason, we can set the lower limit of integration in equation (\ref{eq:asd-rel}) to $r/2$. On the contrary, $a$ can be arbitrarily large and still allow particle periastron to be below any given $r$ (for $e_a(a)\to 1$), contributing to $\asd$ at this radius.

After some manipulation, these considerations allow us to rewrite equation (\ref{eq:asd-rel-gen}) as
\ba  
\asd(r) = \pi^{-1}\int_{r/2}^\infty a^{-1}\Sigma_a(a)\Phi_e(r,a)  \md a,
\label{eq:asd-rel-gen}
\ea  
where 
\ba  
\Phi_e(r,a) =
\frac{\theta(r,a)}{\sqrt{e_a^2(a)-\kappa^2(r,a)}}
\label{eq:Psi_e}
\ea  
is the dimensionless weighting function (a kernel), and 
\ba  
\theta(r,a)=\Theta\left(e_a(a)-\kappa(r,a)\right)
\label{eq:theta1}
\ea  
is equivalent to the definition (\ref{eq:theta}). In Appendix \ref{sec:fluid} we provide an alternative derivation of the relations (\ref{eq:asd-rel-gen})-(\ref{eq:theta1}) based on the results of \citet{Statler01}.

\begin{figure}
	\includegraphics[width=\linewidth]{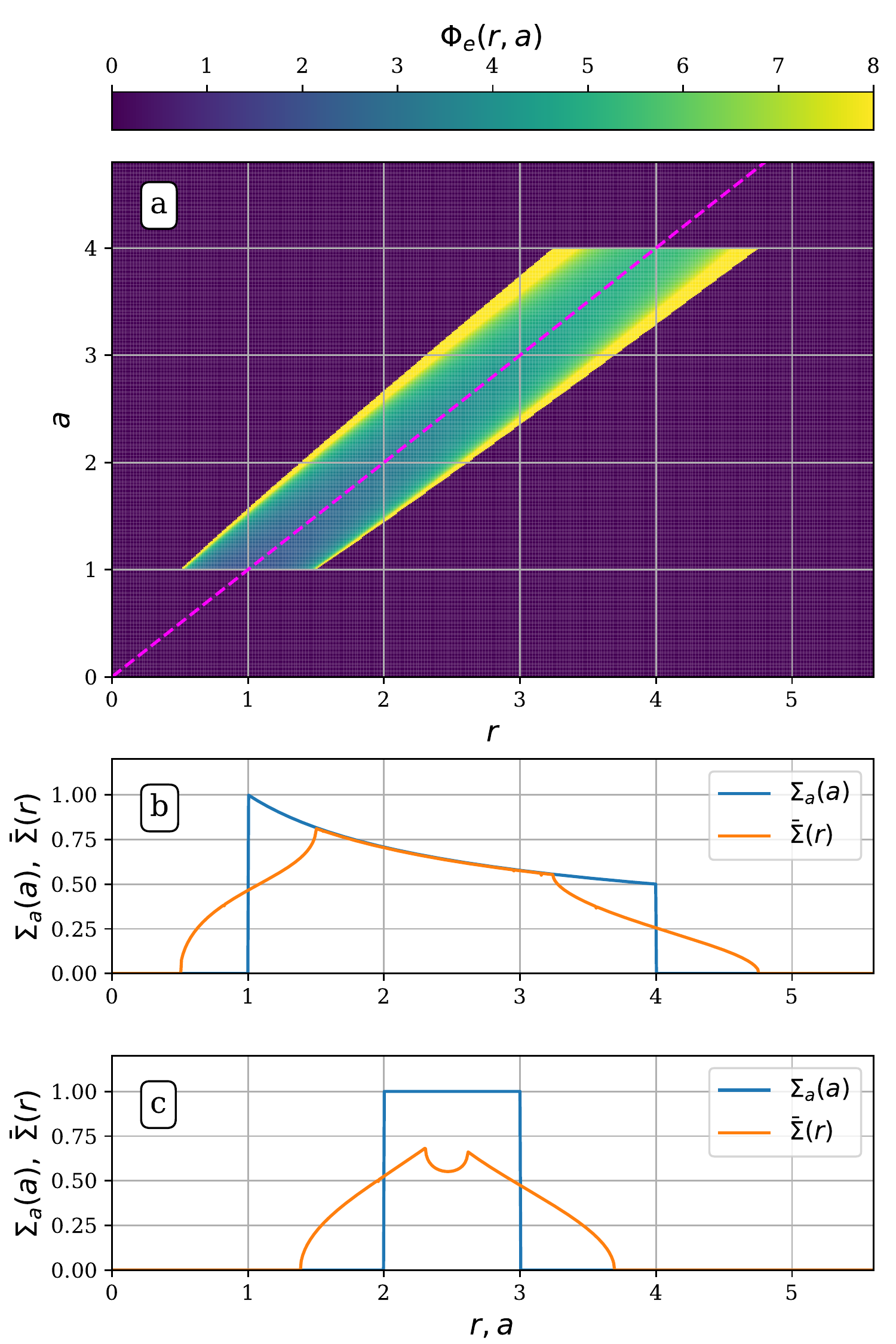}
    \caption{(a) Kernel $\Phi_e(r,a)$ defined by the equation (\ref{eq:Psi_e}) computed as a function of $r$ and $a$ for $e_a(a)$ given by the equation (\ref{eq:e_model}). Note that $\Phi_e(r,a)$ in non-zero for a given $a$ only within a range of $r$ around the purple dashed line $a=r$. (b) Mapping of the distribution of particle semi-major axes $\Sigma_a(a)$ (blue) given by the equation (\ref{eq:Sigma_model1}) into the spatial distribution $\asd(r)$ (orange) using equation (\ref{eq:asd-rel-gen}) and $\Phi_e(r,a)$ from panel (a). (c) Same as in panel (b) but for $\Sigma_a(a)$ given by the equation (\ref{eq:Sigma_model2}).}
    \label{fig:Psi-e}
\end{figure}

Equation (\ref{eq:asd-rel-gen}) is the master equation relating $\Sigma_a(a)$ and $\asd(r)$. Function $\Phi_e(r,a)$ is the key ingredient of this relation. For a given eccentricity profile $e_a(a)$, the kernel $\Phi_e(r,a)$ describes the coupling between the different semi-major axes and the radius $r$, i.e. it determines (through equation (\ref{eq:asd-rel-gen})) how the mass at different $a$ contributes of $\asd$ at $r$. One can easily check that $\Phi_e(r,a)$ satisfies the following integral conditions
\ba  
\int_0^\infty\Phi_e(r,a)dr=\pi a,~~~~\int_0^\infty\Phi_e(r,a)rdr=\pi a^2,
\label{eq:int_constraint}
\ea  
with the second one effectively ensuring mass conservation in the disc when going from $\Sigma_a(a)$ to $\asd(r)$. Note that even though $\Phi_e(r,a)$ formally diverges as $r\to a (1\pm e_a(a))$, these singularities are integrable, resulting in a finite $\asd(r)$ in equation (\ref{eq:asd-rel-gen}). 

Very importantly, in deriving the relations (\ref{eq:asd-rel}) and (\ref{eq:asd-rel-gen})-(\ref{eq:theta1}) we did not use any information about the behavior of apsidal angles of disc particles $\varpi$ --- they can follow any distribution (uniform or non-uniform) without affecting our results. This makes sense since the time that a particle with a given $a$ and $e$ spends within some radial interval is independent of the orientation of its orbital ellipse (which is set by $\varpi$). The insensitivity to apsidal orientation of individual objects comprising the disc allows us to obtain some rather general results, as we show in the rest of the paper.


\subsection{Examples for discs with $e=e_a(a)$}
\label{sec:unique_e-ex}


We illustrate the mapping between $\Sigma_a(a)$ and $\asd(r)$ in Figure \ref{fig:Psi-e}, where in panel (a) we plot $\Phi_e(r,a)$ for an assumed
\ba 
e_a(a)=e_0(a_\mathrm{min}/a)^{0.7},~~~~a_\mathrm{min}<a<a_\mathrm{max},
\label{eq:e_model}
\ea  
with $e_0=0.5$, $a_\mathrm{min}=1$, $a_\mathrm{max}=4$ (in arbitrary units). Note the characteristic structure of $\Phi_e(r,a)$, which is nonzero only for $r_p(a)<r<r_a(a)$ and shows two peaks at any given $a$, near $r_p(a)$ and $r_a(a)$. This behavior is reminiscent of the $\eta(r|a)$ behavior shown in Figure \ref{fig:eta}.

Figure \ref{fig:Psi-e}b,c illustrates the use of this $\Phi_e(r,a)$ to compute $\asd(r)$ using equation (\ref{eq:asd-rel-gen}). In particular, in panel (b) we show how the distribution of particle semi-major axes (blue curve)
\ba 
\Sigma_a(a)=\Sigma_0(a_\mathrm{min}/a)^{1/2},~~~~a_\mathrm{min}<a<a_\mathrm{max},
\label{eq:Sigma_model1}
\ea  
with $\Sigma_0=1$, $a_\mathrm{min}=1$, $a_\mathrm{max}=4$ (in arbitrary units) gets mapped into $\asd(r)$ (orange curve). One can see that $\asd(r)$ is smooth (i.e. continuous), even though $\Sigma_a(a)$ is sharply truncated at the edges (at $a_\mathrm{min}$ and $a_\mathrm{max}$) and $\Phi_e(r,a)$ features (integrable) singularities. Far from the edges $\asd(r)$ is quite similar to $\Sigma_a(a)$ profile, but the sharp edges of $\Sigma_a(a)$ get smeared to a more gentle transition in $\asd(r)$.

Figure \ref{fig:Psi-e}c shows the results of a similar exercise but now for 
\ba
\Sigma_a(a)=\Sigma_0,~~~~a_\mathrm{min}<a<a_\mathrm{max},
\label{eq:Sigma_model2}
\ea
with $\Sigma_0=1$, $a_\mathrm{min}=2$, $a_\mathrm{max}=3$. For this less radially extended $\Sigma_a(a)$, the profile of $\asd(r)$ has a rather different appearance, with two narrowly spaced peaks which arise because of (i) the double-peaked shape of $\Phi_e(r,a)$ at a fixed $a$ and (ii) the proximity of $a_\mathrm{min}$ and $a_\mathrm{max}$ in this case. In particular, $r_p(a_\mathrm{max})<r_a(a_\mathrm{min})$, so that there is an interval of radii where particles with both $a=a_\mathrm{min}$ and $a=a_\mathrm{max}$ are present. 

Note that in constructing $\asd(r)$ in panel (c) we used the same $\Phi_e(r,a)$ shown in panel (a), even though it extends over a larger range of $a$ than $\Sigma_a(a)$ given by the equation (\ref{eq:Sigma_model2}) --- equation (\ref{eq:asd-rel-gen}) naturally takes care of the difference of $(a_\mathrm{min},a_\mathrm{max})$ ranges for $e_a(a)$ and $\Sigma_a(a)$. Thus, a single calculation of $\Phi_e(r,a)$ for a given $e_a(a)$ can be used to generate multiple profiles of $\asd(r)$ for different $\Sigma_a(a)$. Of course, $\Phi_e(r,a)$ should be calculated over the range of $a$ that includes the extent of the assumed $\Sigma_a(a)$, as in Figure \ref{fig:Psi-e}c.


\subsection{Power law discs}
\label{sec:unique_e-pl}


Profiles of $\Sigma_a(a)$ used in making Figure \ref{fig:Psi-e}b,c are of the general form
\ba
\Sigma_a(a)=\Sigma_0(a_\mathrm{min}/a)^{\alpha},~~~~a_\mathrm{min}<a<a_\mathrm{max}.
\label{eq:Sigma_model-pl}
\ea 
For brevity, we will denote them by the set of their parameters, as a model ${\cal M}_\Sigma=(\alpha,\Sigma_0,a_\mathrm{min},a_\mathrm{max})$. For example, in Figure \ref{fig:Psi-e}b we have a model ${\cal M}_\Sigma=(0.5,1,1,4)$, while in Figure \ref{fig:Psi-e}c we show ${\cal M}_\Sigma=(0,1,2,3)$. Similarly, the general power law profile of $e_a(a)$ in the form
\ba 
e_a(a)=e_0(a_0/a)^{\beta}
\label{eq:e_pl}
\ea  
will be denoted as a model ${\cal M}_e=(\beta,e_0,a_0)$
(for reasons outlined in \S \ref{sec:unique_e-ex} we will not worry about specifying the $a$-range for $e(a)$, just assuming that it is large enough), for example, equation (\ref{eq:e_model}) corresponds to ${\cal M}_e=(0.7,0.5,1)$. We will often use such power law models to illustrate our results.

Examination of Figure \ref{fig:Psi-e}b shows that there is a significant range of radii, in which $\asd(r)$ follows the behavior of $\Sigma_a(a)$ rather closely. This correspondence is a rather generic feature of the discs with $\Sigma_a(a)$ and $e_a(a)$ in the power law forms (\ref{eq:Sigma_model-pl}) and (\ref{eq:e_pl}), provided that $r_a(a_\mathrm{min})<r_p(a_\mathrm{max})$ and the relevant range exists. At the same time, it is important to note that the close agreement between $\asd(r)$ and $\Sigma_a(a)$ within this range is not necessarily exact, and in general they differ at some (typically rather small) level. 

We devote Appendix \ref{sec:PL} to a more in-depth exploration of these characteristics of the power law discs, highlighting the points made above.
 

\section{disc with a distribution of eccentricity}
\label{sec:e_dist}


Next we consider a more general situation, in which particles at a given semi-major axis $a$ might have a {\it distribution} of eccentricities $\psi_e(e,a)$, with properties (e.g. mean, dispersion, etc.) that also depend on $a$. We will assume that $\int_0^1 \psi_e(e,a)\md e=1$ at every $a$. 

Generalizing the results of \S\ref{sec:unique_e}, we introduce a function $\eta(r|a,e)$ such that $\eta(r|a,e)\md r\,\md a\,\md e$ is the amount of mass per radial interval $(r,r+\md r)$ contributed by objects with eccentricity in the interval $(e,e+\md e)$ and semi-major axes in the interval $(a,a+\md a)$. Clearly, $\eta(r|a,e)$ is non-zero only if the condition (\ref{eq:cond}) is fulfilled, in which now $r_p=r_p(a,e)$, $r_a=r_a(a,e)$. The normalization condition (\ref{eq:norm}) generalizes to
\ba
\md a\,\md e\int_{r_p(a)}^{r_a(a)}\eta(r|a,e)\md r=\psi_e(e,a)\,\md m(a)\,\md e.
\label{eq:norm-e}
\ea  
Repeating the development in \S\ref{sec:unique_e}, and extending equation (\ref{eq:etar}) by adding integration over eccentricity, we obtain
\ba  
\asd(r)=\pi^{-1}\int_0^\infty\md a \int_0^1\md e\frac{\Sigma_a(a)\psi_e(e,a)\theta(r,e,a)}{\sqrt{(r-a(1-e))(a(1+e)-r)}},
\label{eq:asd-rel-gen0}
\ea  
which is a natural generalization of the equation (\ref{eq:asd-rel}). The generalized indicator function 
\ba  
\theta(r,a,e)=\Theta\left([r-a(1-e)][a(1+e)-r]\right),
\label{eq:theta-gen}
\ea  
ensures that the constraint (\ref{eq:cond}) is obeyed when integration over $e$ and $a$ is carried out. That constraint can also be written as $e>\kappa(r,a)$, analogous to the equation (\ref{eq:kappa}).

Finally, by manipulating equation (\ref{eq:asd-rel-gen0}) one can again put it in the form (\ref{eq:asd-rel-gen}) but with a new weighting function $\Phi_e$ given by 
\ba  
\Phi_e(r,a) =
\int_{\kappa}^1 \frac{\psi_e(e,a)}{\sqrt{e^2-\kappa^2}} \md e
\label{eq:Psi_gen}
\ea  
with $\kappa=\kappa(r,a)$. It is easy to see that the integral in (\ref{eq:Psi_gen}) is convergent so that $\Phi_e$ is finite.

Equation (\ref{eq:Psi_gen}) represents a natural generalization of the equation (\ref{eq:Psi_e}) in \S\ref{sec:unique_e} to the case of a distribution of eccentricities (at any $a$). For $\psi_e(e,a)$ in the form (\ref{eq:Dirac}) the expression (\ref{eq:Psi_gen}) for $\Phi_e$ naturally reduces to the equation (\ref{eq:Psi_e}). One can easily check that the relations (\ref{eq:int_constraint}) still hold for $\Phi_e$ in the form (\ref{eq:Psi_gen}). Also, according to definitions (\ref{eq:kappa}) and (\ref{eq:Psi_gen}), $\Phi_e(r,a)$ must obey the following symmetry property:
\ba  
\Phi_e(a-z,a)=\Phi_e(a+z,a)
\label{eq:symmetry}
\ea  
for any $a$ and $z$. 

The weighting function $\Phi_e$ absorbs all information about the shape of the eccentricity distribution $\psi_e(e,a)$, with equation (\ref{eq:Psi_gen}) providing a unique correspondence between the two (we will show in \S\ref{sec:inverse-e} that this relation can be inverted). In Appendix \ref{sec:psi-rings} we provide examples of such $\psi_e-\Phi_e$ pairs, which can be used in applications.  

Equations (\ref{eq:asd-rel-gen}) and (\ref{eq:Psi_gen}) represent the desired relation of the ASD $\asd$ to the mass distribution in the disc (i.e. $\Sigma_a(a)$) and its kinematic properties (i.e. $\psi_e(e,a)$). They allow efficient computation of $\asd(r)$ once $\Sigma_a(a)$ and $\psi_e(e,a)$ are specified. 

\begin{figure}
	\includegraphics[width=\linewidth]{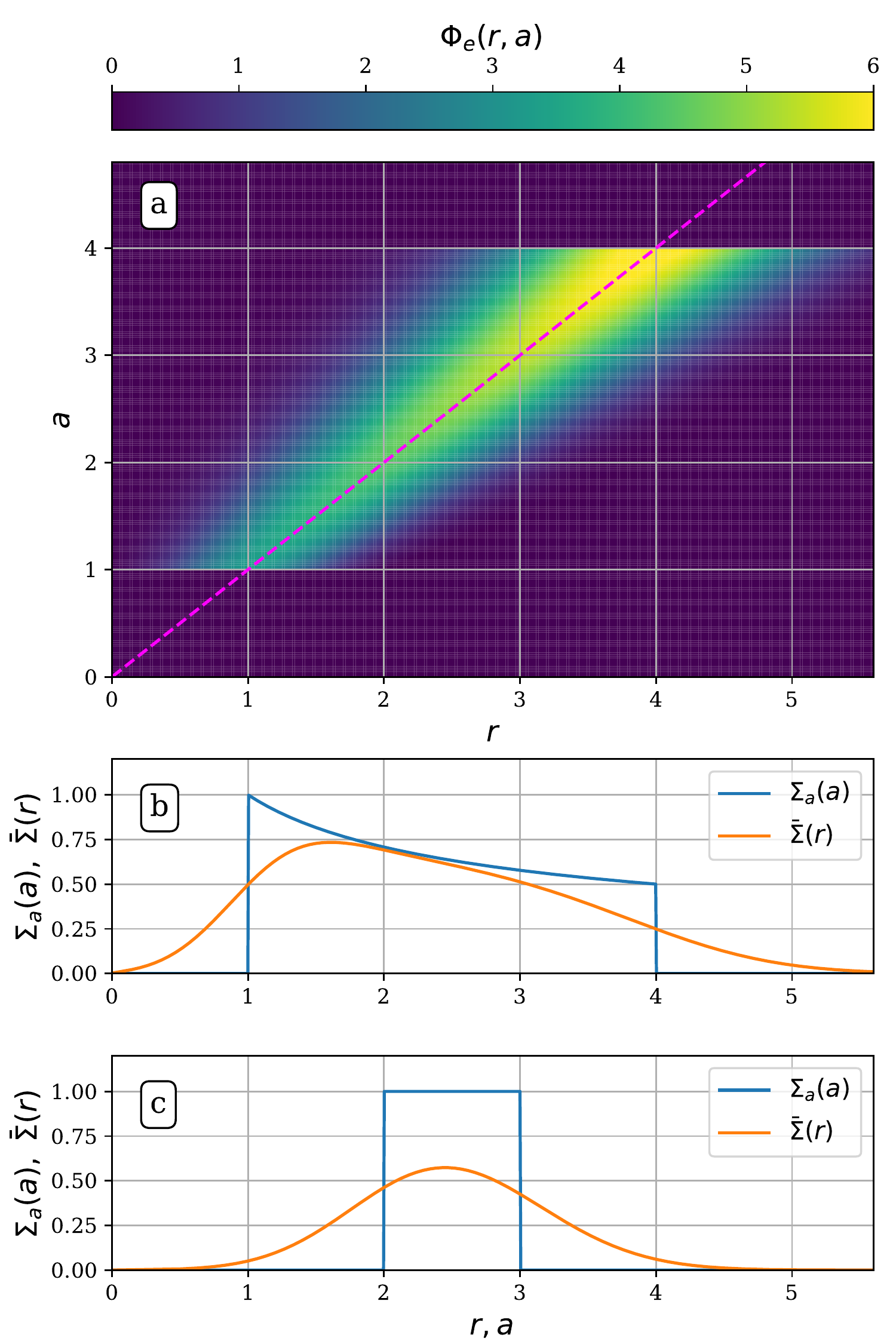}
    \caption{Same as Figure \ref{fig:Psi-e}, but for a (truncated) Rayleigh distribution (\ref{eq:Rayleigh}) of the debris particle eccentricities. (a) Kernel $\Phi_e(r,a)$ given by the equation (\ref{eq:Psi_Ray}) with $\sigma(a)$ in the form (\ref{eq:e_dis_model}). (b) Mapping between $\Sigma_a(a)$ (blue) to $\asd(r)$ (orange) using this $\Phi_e(r,a)$ and equation (\ref{eq:asd-rel-gen}). This calculation assumes $\Sigma_a(a)$ in the form (\ref{eq:Sigma_model1}). (c) Same as panel (b) but for $\Sigma_a(a)$ in the form (\ref{eq:Sigma_model2}).}
    \label{fig:Psi-e-Gauss}
\end{figure}


\subsection{Rayleigh distribution of eccentricity}
\label{sec:Rayleigh}


We now illustrate our results for a disc with the Rayleigh distribution of eccentricity, which is expected to be established as a result of two-body relaxation (self-stirring) in the disc \citep{Ida1992,KenW2010}. Namely, we will consider $\psi_e(e,a)$ in the form 
\ba  
\psi_e(e,a)=\frac{e}{\sigma^2(1-e^{-1/2\sigma^2})}e^{-e^2/2\sigma^2},
\label{eq:Rayleigh}
\ea  
where $\sigma$ is the dispersion of eccentricity, which can be a function of $a$ in general. Note that the factor multiplying the exponential in (\ref{eq:Rayleigh}) accounts for the fact that $e<1$ and the normalization $\int_0^1 \psi_e(e,a)\md e=1$.

Plugging (\ref{eq:Rayleigh}) in the equation (\ref{eq:Psi_gen}) and integrating we find 
\ba  
\Phi_e(r,a) = \left(\frac{\pi}{2}\right)^{1/2}\frac{e^{-\kappa^2/2\sigma^2}}{\sigma(1-e^{-1/2\sigma^2})} {\rm erf} \left(\sqrt{\frac{1-\kappa^2}{2\sigma^2}}\right),
\label{eq:Psi_Ray}
\ea  
where $\sigma=\sigma(a)$, $\kappa=\kappa(r,a)$ as per equation (\ref{eq:kappa}), and ${\rm erf}(z) = 2\pi^{-1/2}\int_0^z e^{-t^2}\md t$ is the Gauss error function. Relation (\ref{eq:asd-rel-gen}) between $\Sigma_a(a)$ and $\asd(r)$ can then be written as
\ba  
\asd(r) = \frac{1}{\sqrt{2\pi}}\int_{r/2}^\infty \frac{\Sigma_a(a)e^{-\kappa^2/2\sigma^2}}{a\sigma (1-e^{-1/2\sigma^2})} {\rm erf} \left(\sqrt{\frac{1-\kappa^2}{2\sigma^2}}\right)  \md a,
\label{eq:asd-rel-Ray}
\ea  
which is a convolution of $\Sigma_a(a)$ with a known kernel. These results are valid for arbitrary eccentricity dispersion, including $\sigma\sim 1$.


\subsection{Examples for discs with the Rayleigh eccentricity distribution}
\label{sec:Gauss_e-ex}


In Figure \ref{fig:Psi-e-Gauss}a we plot $\Phi_e(r,a)$ given by the equation (\ref{eq:Psi_Ray}), assuming that  
\ba 
\sigma(a)=\sigma_0(a_\mathrm{min}/a)^{1/2},~~~~a_\mathrm{min}<a<a_\mathrm{max},
\label{eq:e_dis_model}
\ea  
with $\sigma_0=0.4$, $a_\mathrm{min}=1$, $a_\mathrm{max}=4$ (in arbitrary units). Note that, unlike Figure \ref{fig:Psi-e}a, $\Phi_e(r,a)$ for the Rayleigh eccentricity distribution has a single peak around $r=a$ line, and exhibits smooth decay away from it. 

In panels (b) and (c) we show $\asd(r)$ computed for the same $\Sigma_a(a)$ profiles as in Figure \ref{fig:Psi-e}b,c, respectively (i.e. ${\cal M}_\Sigma=(0.5,1,1,4)$ and ${\cal M}_\Sigma=(0,1,2,3)$), but now using the Rayleigh $\Phi_e(r,a)$ from Figure \ref{fig:Psi-e-Gauss}a. One can see that $\asd(r)$ obtained for the Rayleigh distribution of eccentricities is smooth and does not show sharp changes in its slope even near $a_\mathrm{min}$ and $a_\mathrm{max}$, unlike the ASD in Figure \ref{fig:Psi-e}b,c. Also, even far from the edges of $\Sigma_a(a)$ profiles, the behavior of $\asd(r)$ is not as close to $\Sigma_a(r)$ as in Figure \ref{fig:Psi-e}, c.f. \S\ref{sec:unique_e-pl} (they would follow each other better for smaller $\sigma(a)$).

\begin{figure}
	\includegraphics[width=\linewidth]{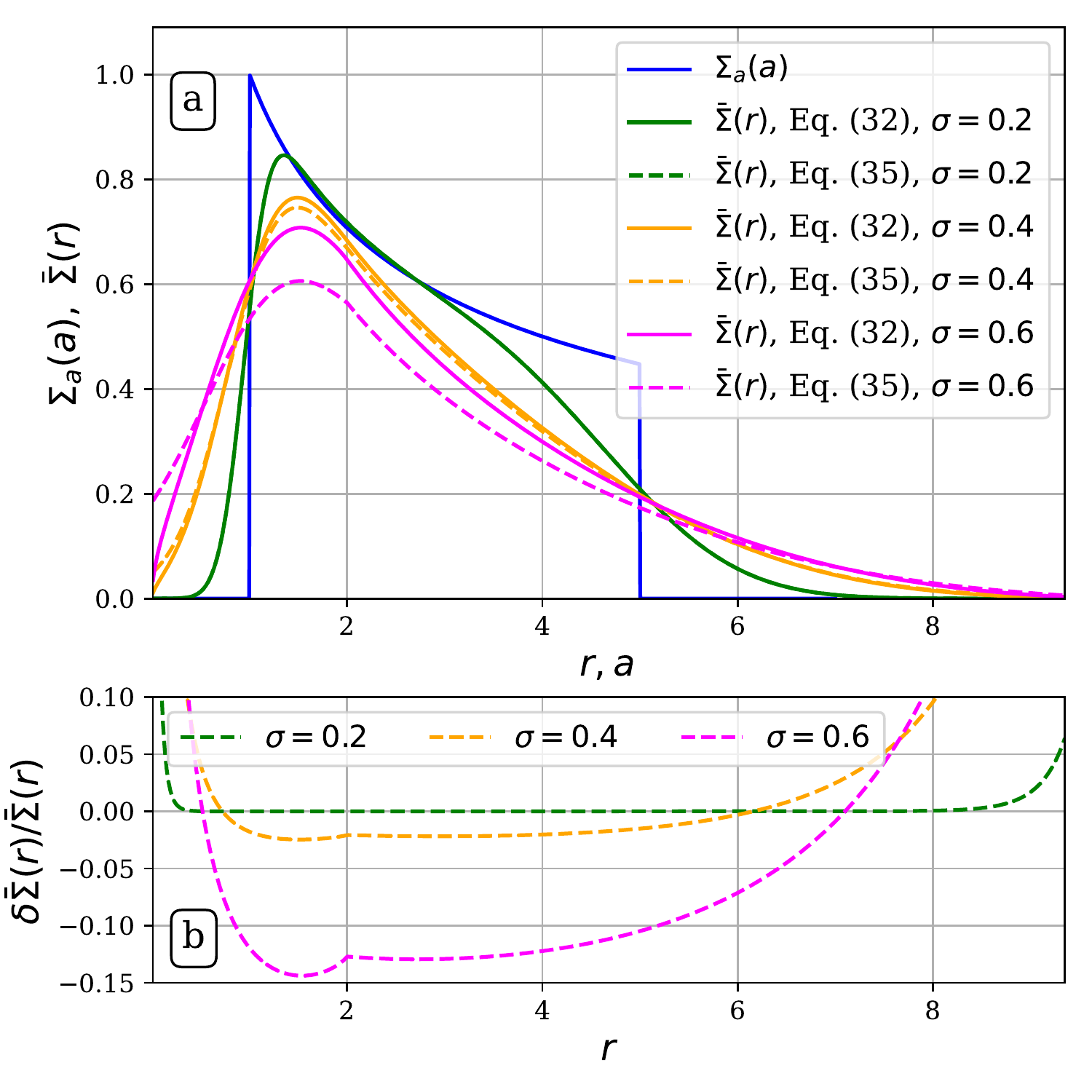}
    \caption{Illustration of the accuracy of the small-$e$ approximation (\ref{eq:Psi_Ray_app})-(\ref{eq:asd-rel-Ray_app}) for a (truncated) Rayleigh eccentricity distribution, using power law disc model ${\cal M}_\Sigma=(0.5,1,1,5)$ for $\Sigma_a(a)$ (blue curve). (a) Profiles of $\asd(r)$ for three values of eccentricity dispersion $\sigma=0.2$ (green), $\sigma=0.4$ (orange) and $\sigma=0.6$ (magenta) computed using equations (\ref{eq:Psi_Ray})-(\ref{eq:asd-rel-Ray}) (solid) and (\ref{eq:Psi_Ray_app})-(\ref{eq:asd-rel-Ray_app}) (dashed). (b) Relative deviation between the exact calculations and low-$e$ approximations for $\asd(r)$ shown in panel (a).}
    \label{fig:small-e}
\end{figure}


\subsection{Small eccentricity limit, $\sigma(a)\ll 1$, for the Rayleigh distribution}
\label{sec:low_e}


Of large practical significance is the situation when the eccentricities of debris disc particles are small, $e\ll 1$, meaning that $\sigma(a)\ll 1$. In this limit $\Phi_e(r,a)$ is non-negligible only for $\kappa\lesssim \sigma$ because of the exponential term in (\ref{eq:Psi_Ray}). Physically this means that particles with semi-major axes $|r-a|\lesssim \sigma r$ (i.e. a typical radial excursion of disc particles for $\psi_e$ in the form (\ref{eq:Rayleigh})) provide the dominant contribution to ASD at $r$. For $\kappa\lesssim \sigma\ll 1$ the argument of ${\rm erf}$ is of order $\sigma^{-1}\gg 1$, so that ${\rm erf}\to 1$. Also, $e^{-1/2\sigma^2}\to 0$. As a result, for $\sigma\ll 1$ we can reasonably approximate 
\ba  
\Phi_e(r,a) &\approx & \left(\frac{\pi}{2}\right)^{1/2}\sigma^{-1}
\exp\left(-\frac{\kappa^2}{2\sigma^2}\right),
\label{eq:Psi_Ray_app}\\
\asd(r) &\approx & \frac{1}{\sqrt{2\pi}}\int_{r/2}^\infty \frac{\Sigma_a(a)}{a\sigma} \exp\left(-\frac{\kappa^2}{2\sigma^2}\right)  \md a.
\label{eq:asd-rel-Ray_app}
\ea  

In Figure \ref{fig:small-e} we illustrate how well the approximations (\ref{eq:Psi_Ray_app}), (\ref{eq:asd-rel-Ray_app}) work compared to the exact equations (\ref{eq:Psi_Ray}), (\ref{eq:asd-rel-Ray}) for different values of (spatially constant) $\sigma$ using a power law disc model (see \S\ref{sec:unique_e-pl}) ${\cal M}_\Sigma=(0.5,1,1,5)$. One can see that even for $\sigma$ as high as $0.4$ the difference between the exact and approximate calculations of $\asd(r)$ is below 3$\%$ for $a_\mathrm{min}\lesssim a\lesssim a_\mathrm{max}$; the two methods are essentially indistinguishable for $\sigma=0.2$. Only for $\sigma=0.6$ one finds deviations at the level of $15\%$. Thus, the approximations (\ref{eq:Psi_Ray_app}), (\ref{eq:asd-rel-Ray_app}) perform very well, even for quite substantial values of $\sigma$.


\section{Reconstruction of mass distribution from observations}
\label{sec:inverse-mass}


Observations of debris discs allow us to directly obtain their ASD $\asd(r)$. A natural question to ask then is whether this information can be used to infer the distribution of particle semi-major axes $\Sigma_a(a)$ and/or the eccentricity distribution $\psi_e(e,a)$. In other words, one would like to use observations to obtain the underlying disc properties by somehow inverting the equation (\ref{eq:asd-rel-gen}).

It is clear that the knowledge of $\asd(r)$ cannot help us uniquely determine both $\Sigma_a(a)$ and $\psi_e(e,a)$ at the same time: multiple pairs of $\Sigma_a(a)$ and $\Phi_e(e,a)$ can result in the same $\asd(r)$ via the equation (\ref{eq:asd-rel-gen}). Thus, some additional information about one of these functions has to be provided based on some additional considerations. In this section we will assume that we somehow possess independent knowledge of $\psi_e(r,a)$. 

For example, as already alluded to in \S\ref{sec:Rayleigh}, particle discs dynamically excited by self-stirring are expected to follow Rayleigh distribution of eccentricities $\psi_e$ \citep{Ida1992}. Moreover, the dispersion of particle inclinations in such discs is typically found to be about a half of the local eccentricity dispersion \citep{Ida1992,IdaKM1993}. If a self-stirred debris disc is observed edge-on (relative to our line of sight), the measurement of the radial variation of its vertical thickness would provide us with the radial profile of the particle inclinations $i(a)$, or the dispersion of inclinations, see \citet{Han2022}. Consequently, the eccentricity dispersion as a function of $a$ can be obtained as $\sigma(a)\approx 2i(a)$, giving us the full information about $\psi_e(e,a)$. 

Another situation in which we could have a priori knowledge of particle dynamics arises when particle eccentricities are known to be secularly excited by a massive planet\footnote{Close to being co-planar with the disc to not violate our assumption of a two-dimensional disc.}. In this case an assumption of a unique $e=e_a(a)$ (as in \S\ref{sec:unique_e}) would be natural. Moreover, if the age of the system, as well as the planetary mass and orbital properties (i.e. its semi-major axis and eccentricity) are known, then one could use the classical Laplace-Lagrange secular theory \citep{Murray1999} to predict the profile of $e_a(a)$ in the disc.

In both situations we would know the behavior of $\psi_e(e,a)$, allowing us to compute $\Phi_e(r,a)$ via equations (\ref{eq:Psi_e}) or (\ref{eq:Psi_gen}), as appropriate. Then, by making a change of variable $r\to 2x$ we can formally convert the relation (\ref{eq:asd-rel-gen}) to a linear Volterra integral equation of the first kind
\ba 
\asd(2x)=\int_x^\infty K(x,a)\Sigma_a(a)\md a,
\label{eq:volt}
\ea 
with the kernel
\ba  
K(x,a)=\frac{\Phi_e(2x,a)}{\pi a}.
\label{eq:kernel}
\ea 
This equation can be discretized on a grid, converting it to a linear (upper triangular) matrix equation, which can be easily solved for $\Sigma_a(a)$ using standard inversion methods (e.g. forward substitution, \citealt{NR2002}), since $\asd(2x)$ is available from observations and $K(x,a)$ is known (as assumed). 

Thus, knowledge of the dynamical characteristics of a debris disc allows one to use its observed $\asd(r)$ to infer the semi-major axis distribution $\Sigma_a(a)$ of particles in the disc in a straightforward manner.


\section{Reconstruction of eccentricity distribution from observations}
\label{sec:inverse-e}


We will now look into the question of whether the knowledge of ASD $\asd(r)$ can be used to retrieve the distribution of particle eccentricities $\psi_e(e,a)$.

We start by noting that equation (\ref{eq:Psi_gen}) can be viewed as a version of the Abel integral equation for the unknown function $\psi_e(e,a)$ \citep{Binney2008}, with $a$ being a parameter. Also, we can consider $\Phi_e$ to be a function of $\kappa$ and $a$ (instead of $r$ and $a$) as $\Phi_e(a(1\pm\kappa),a)$, see equation (\ref{eq:kappa}). The sign ambiguity of $\kappa$ is not important because of the property (\ref{eq:symmetry}) which $\Phi_e$ must obey in the first place. Then equation (\ref{eq:Psi_gen}) can be easily inverted as\footnote{This solution can be checked by substituting the definition (\ref{eq:Psi_gen}) for $\Phi_e$, differentiating with respect to $\kappa$, and changing the order of integration in the resultant double integral. To show that $\Phi_e$ in the form (\ref{eq:Psi_e}) leads to $\psi_e$ in the form (\ref{eq:Dirac}), one should first re-write it as  $\Phi_e=\int_\kappa^1\delta(e-e_a(a))(e^2-\kappa^2)^{-1/2}\md e$ and then carry out the above procedure.} \citep{Binney2008} 
\ba  
\psi_e(e,a)=-\frac{2e}{\pi}\int_e^1
\frac{d\kappa}{\sqrt{\kappa^2-e^2}}\frac{\partial}{\partial\kappa}\Phi_e(a(1\pm \kappa),a).
\label{eq:Abel}
\ea  
Thus, if the function $\Phi_e(r,a)$ defined by the equation (\ref{eq:Psi_gen}) is somehow known to us as a function of $r$ and $a$, then the behavior of $\psi_e(e,a)$ can be determined uniquely through the equation (\ref{eq:Abel}). But the really tricky part of this exercise is obtaining the knowledge of $\Phi_e(r,a)$ behavior in the first place. 

Indeed, let us assume that we know exactly the behavior of both $\asd(r)$ and $\Sigma_a(a)$ in the whole disc. Even then we would still not be able to infer $\Phi_e(r,a)$ from the equation (\ref{eq:asd-rel-gen}): mathematically, this is an ill-posed problem equivalent to inferring a kernel of an integral equation and has no unique solution\footnote{Discretization of the equation (\ref{eq:asd-rel-gen}) on $N\times N$ grid in $(r,a)$ space yields $N$ equations for $N^2$ unknown components of the discretized kernel $\Phi_e$.}. This is unlike the situation considered in \S\ref{sec:inverse-mass}, where we have shown that the knowledge of the kinematic properties of the disc (i.e. $\psi_e(e,a)$) allows us to infer $\Sigma_a(a)$.

Progress can still be made if the functional form of the eccentricity distribution is known, and we are just looking to determine the dependence of one of its characteristics as a function of $a$. For example, if for whatever reason we expect particle eccentricity to be a unique function of $a$, as in \S\ref{sec:unique_e} (i.e. $\psi_e$ to have the form (\ref{eq:Dirac})), then equations (\ref{eq:asd-rel-gen}), (\ref{eq:Psi_e}) can be combined to provide an integral equation for $e_a(a)$ only (once $\asd(r)$ and $\Sigma_a(a)$ are specified). As a complicating factor, $e_a(a)$ enters the equation (\ref{eq:Psi_e}) in a {\it nonlinear} fashion, unlike $\Sigma_a$ in equation (\ref{eq:volt}). However, the original integral equation (\ref{eq:asd-rel-gen}), (\ref{eq:Psi_e}) can still be reduced to a (nonlinear) Volterra equation of the first kind by the same method as used in \S\ref{sec:inverse-mass} for equation (\ref{eq:volt}). The numerical solution of such an equation, even if nonlinear, is still reasonably straightforward, see \citet{NR2002}.  

A similar situation arises for self-stirred discs, in which we expect $\psi_e(e,a)$ to be given by the Rayleigh distribution (\ref{eq:Rayleigh}). In this case we would like to determine the run of $\sigma(a)$. As we have shown in \S\ref{sec:Rayleigh}, for such discs equation (\ref{eq:asd-rel-Ray}) provides an integral relation between the known $\asd(r)$, $\Sigma_a(a)$ and unknown $\sigma(a)$. Again, $\sigma(a)$ enters this equation in a highly non-linear fashion, but the same method based on a reduction of (\ref{eq:asd-rel-Ray}) to a nonlinear Volterra equation would still apply for the determination of $\sigma(a)$.

Also, in some special cases one can determine $\psi_e(e,a)$ in a non-parametric form (i.e. not assuming a particular functional dependence on $e$). This possibility arises for certain special forms of $\Sigma_a(a)$ and is discussed next.


\subsection{Eccentricity distribution from observations of narrow rings}
\label{sec:rings-e}

\begin{figure}
	\includegraphics[width=\linewidth]{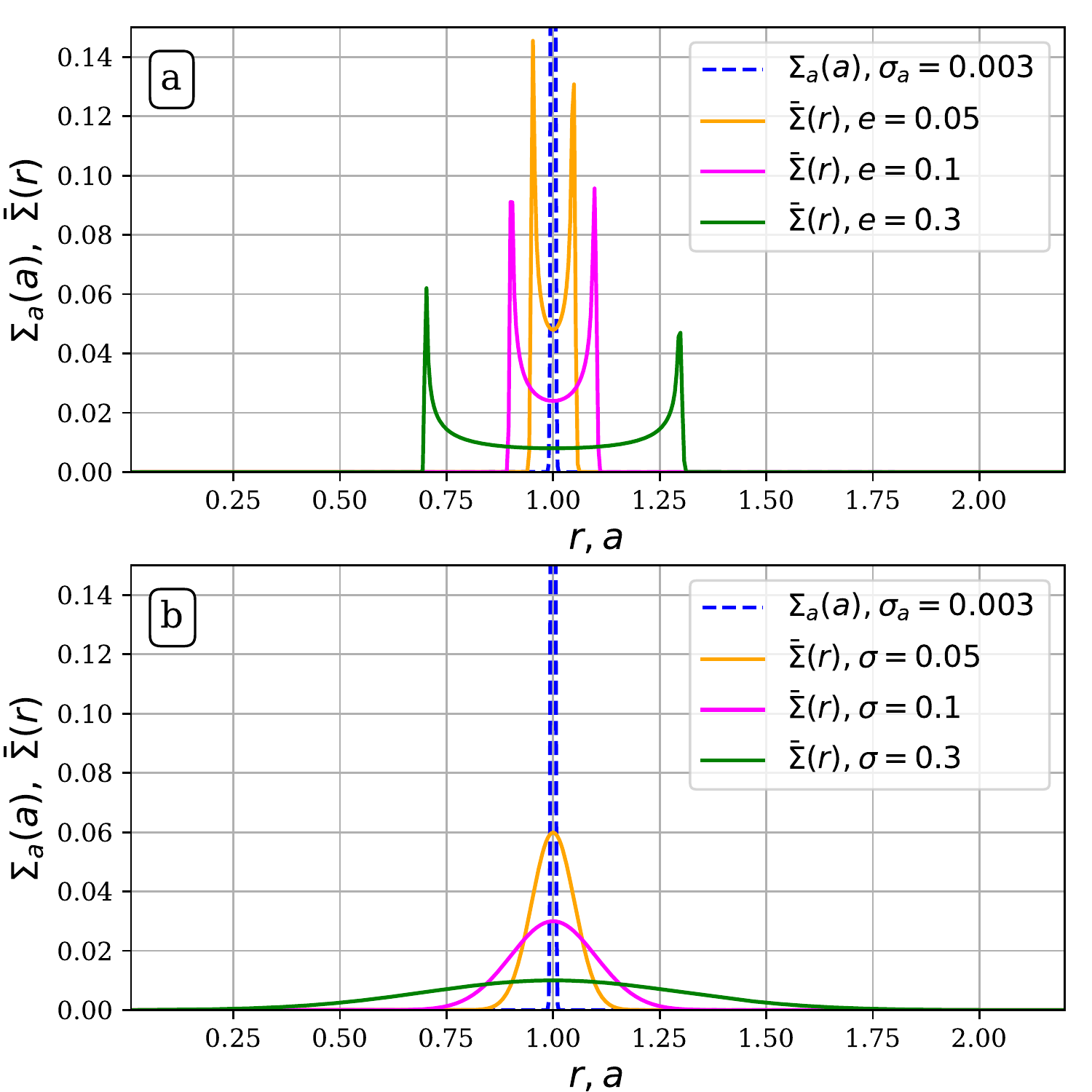}
    \caption{Profiles of $\asd(r)$ produced by a narrow (in semi-major axis) ring. This calculation assumes $\Sigma_a(a)$ to be a Gaussian with a small width $\sigma_a=0.003$ (dashed blue curve, approximating the Eq. (\ref{eq:ring})). (a) Calculation for several values of $e_a$, assuming a unique value of eccentricity. (b) Calculation for a Rayleigh eccentricity distribution for several values of $\sigma$. }
    \label{fig:ring}
\end{figure}

Let us assume that it is known a priori that the debris particles are clustered in semi-major axis around $a=a_r$ such that their relative spread in semi-major axes, $|a-a_r|/a_r$, is much smaller than their characteristic eccentricity (i.e. its mean or dispersion). This situation is illustrated in Figure \ref{fig:ring}, where we show the characteristic runs of $\asd(r)$ for a ring with a fixed $e_a$ (computed using equation (\ref{eq:Psi_e}), for three values of $e_a$), and for a ring with a Rayleigh distribution of eccentricities (obtained using equation (\ref{eq:Psi_Ray}), for three values of $\sigma$). While the underlying $\Sigma_a(a)$ in the form of a narrow Gaussian (with the dispersion of $0.003$ in $a$) is the same, the ASDs differ drastically in two cases, e.g. double-peaked in (a) vs. single peaked in (b), easily revealing the difference between the underlying eccentricity distributions \citep{Kennedy2020}.  

More generally, we can represent a narrow ring (of mass $m_r$) by assuming that all its particles have the same semi-major axis $a_r$, such that
\ba  
\Sigma_a(a)=\frac{m_r}{2\pi a_r} \delta(a-a_r).
\label{eq:ring}
\ea
Plugging this $\Sigma_a(a)$ into the equation (\ref{eq:asd-rel-gen}), one finds
\ba 
\asd(r)=\frac{m_r}{2\pi^2 a_r^2}\Phi_e(r,a_r).
\label{eq:Sig-ring}
\ea 
Thus, for a narrow ring we can use the knowledge of the radial profile of $\asd(r)$ to directly obtain the dependence of $\Phi_e$ on $r$:
\ba 
\Phi_e(r,a_r)=\frac{2\pi^2 a_r^2}{m_r}\asd(r).
\label{eq:Psi-ring}
\ea 
Plugging this $\Phi_e$ in the equation (\ref{eq:Abel}) and setting $a=a_r$ one can subsequently determine the explicit dependence of $\psi_e(e,a_r)$ on $e$. 

In Appendix \ref{sec:psi-rings} we illustrate this $\psi_e$-reconstruction technique using a variety of ring ASD profiles. In the process, we also derive a number of $\psi_e$--$\Phi_e$ pairs mutually related via the equations (\ref{eq:Psi_gen}) and (\ref{eq:Abel}), which may be useful in application to observations.

Note that the assumption of an underlying ring-like $\Sigma_a(a)$ can be justified only if $\asd$ is such that 
\ba  
\asd(a_r+z)=\asd(a_r-z)
\label{eq:sym-ring}
\ea  
for any $z$, which immediately follows from equations (\ref{eq:symmetry}) and (\ref{eq:Psi-ring}). If this constraint is not fulfilled then the debris disc particles cannot be strongly clustered in semi-major axis. But if (\ref{eq:sym-ring}) is fulfilled, then the narrow ring interpretation with a characteristic ring width $\Delta a_r\ll ea_r$ (or $\sigma a_r$) is quite plausible. 

Examination of Figure \ref{fig:ring} shows that constraint (\ref{eq:sym-ring}) is fulfilled for a Gaussian ring (panel (b)) with a very high accuracy. However, a ring with a fixed $e$ (panel (a)) reveals noticeable asymmetries between the inner and outer peaks of the ASD. It turns out that these asymmetries arise from (i) the finite (i.e. non-zero, even if rather small) width of the ring in this example, and (ii) the divergent nature of $\Phi_e(r,a_r)$ in the form (\ref{eq:Psi_e}) near the periastra and apoastra of the ring particles. The amplitude of these asymmetries may be used to measure the radial width $\Delta a_r$ of the semi-major axis distribution of the ring particles.


\subsection{Eccentricity distribution from observations of sharp disc edges}
\label{sec:edges-e}


Another possibility to constrain $\psi_e$ arises at the sharp edges of $\Sigma_a(a)$ distribution of debris discs particles. We will make two assumptions. First, that both $\psi_e(e,a)$ and $\Sigma_a(a)$ vary with $a$ slowly inside the disc near its edge at $a_0$, i.e. on scales larger than the typical epicyclic excursion $ea_0$ of particles near the edge. Second, that the drop of $\Sigma_a$ at the edge occurs over the range of $a$, which is much smaller than the typical epicyclic excursion $ea_0$. Once these conditions are fulfilled, we can (i) assume that $\psi_e(e,a)$ is independent of $a$ near the edge and (ii) adopt $\Sigma_a(a)$ in the form (without loss of generality we will focus on the outer disc edge)  
\ba 
\Sigma_a(a)=\Sigma_0\Theta(a_0-a),
\label{eq:Sig-edge}
\ea 
i.e. $\Sigma_a(a)=\Sigma_0$ for $a<a_0$ and zero outside $a_0$. 

We will now illustrate the $\asd(r)$ behavior near the disc edge for two different assumptions about disc kinematics (\S\ref{sec:edges-ea},\ref{sec:edges-Ray}) focusing on the low-$e$ case, and then provide a general recipe for $\Phi_e$ retrieval near the disc edge (\S\ref{sec:edges-gen}). 


\subsubsection{$\asd(r)$ at the edge of a disc with $e=e_a(a)$ }
\label{sec:edges-ea}

\begin{figure}
	\includegraphics[width=\linewidth]{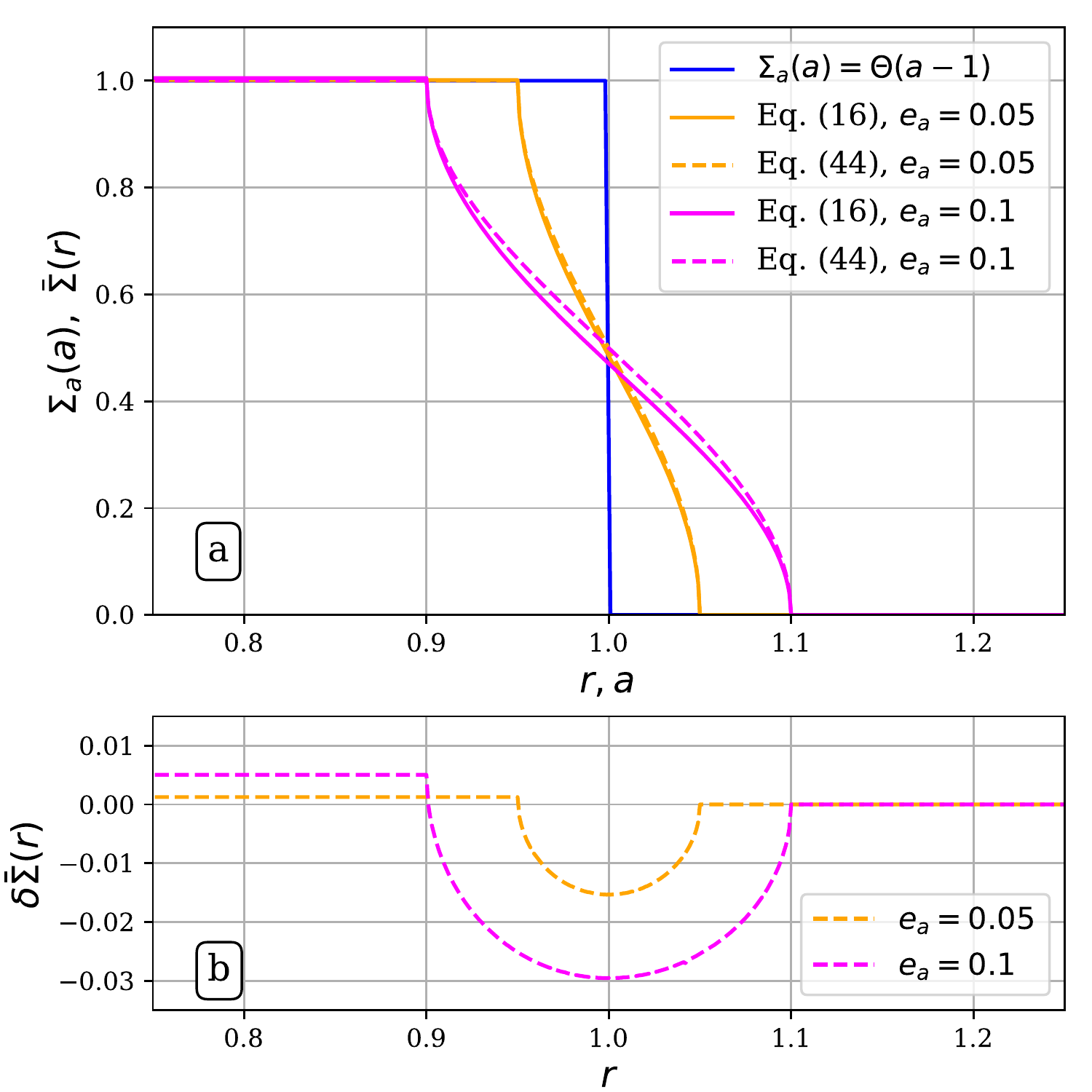}
    \caption{(a) Profiles of $\asd(r)$ near a sharp edge of $\Sigma_a(a)$ (at $a_0=1$) for a disc with a fixed value of $e=e_a$ near the edge. Calculations are performed for $e_a=0.05$ (orange) and $0.1$ (purple), assuming $\Sigma_a(a)=1$ for $a<1$ and zero otherwise. Solid curves are computed using Eqs. (\ref{eq:asd-rel-gen})-(\ref{eq:theta1}), while the dashed curves represent the small-$e$ approximation (\ref{eq:Sig-edge-e}). (b) The difference between the exact and approximate (i.e. Eq. (\ref{eq:Sig-edge-e})) calculations of $\asd(r)$ shown in panel (a).}
    \label{fig:edge-e-a}
\end{figure}

In Figure \ref{fig:edge-e-a} we show the profile of $\asd(r)$ near the edge of a disc with $e=e_a$, which is independent of $a$. This calculation (solid curves) is done for two values of $e_a$ and uses equations (\ref{eq:asd-rel-gen})-(\ref{eq:Psi_e}).

One can also show using the same equations (\ref{eq:asd-rel-gen})-(\ref{eq:Psi_e}) that in the low-$e$ limit a disc with $e=e_a$ and a sharply truncated edge has $\asd(r)=\Sigma_0$ for $r \le r_p$, while
\ba  
\asd(r)=\frac{\Sigma_0}{2}\left[1-\frac{2}{\pi}{\rm asin}\frac{r-a_0}{ea_0}\right], ~~~\mbox{for}~~~r_p\le r\le r_a,
\label{eq:Sig-edge-e}
\ea  
and $\asd(r)=0$ for $r \ge r_a$, where $r_p=a_0(1-e_a)$ and $r_a=a_0(1+e_a)$. 

We show this approximation for $\asd(r)$ in Figure \ref{fig:edge-e-a}a (dashed lines) for the same two values of $e_a$ as the exact calculation, and their difference in panel (b). One can see that small differences (up to $3\%$ for $e_a=0.1$) arise in the near-edge region $r_p\le r\le r_a$. Also, for $r<r_p$ the exact $\asd(r)$ is constant but slightly higher than $\Sigma_0$ with the difference rapidly diminishing with decreasing $e_a$. This behavior is explained in Appendix \ref{sec:PL}. Aside from these differences, the approximation (\ref{eq:Sig-edge-e}) provides a reasonable match to $\asd(r)$, especially at low $e_a$.

\subsubsection{$\asd(r)$ at the edge of a disc with a Rayleigh $\psi_e$}
\label{sec:edges-Ray}

\begin{figure}
	\includegraphics[width=\linewidth]{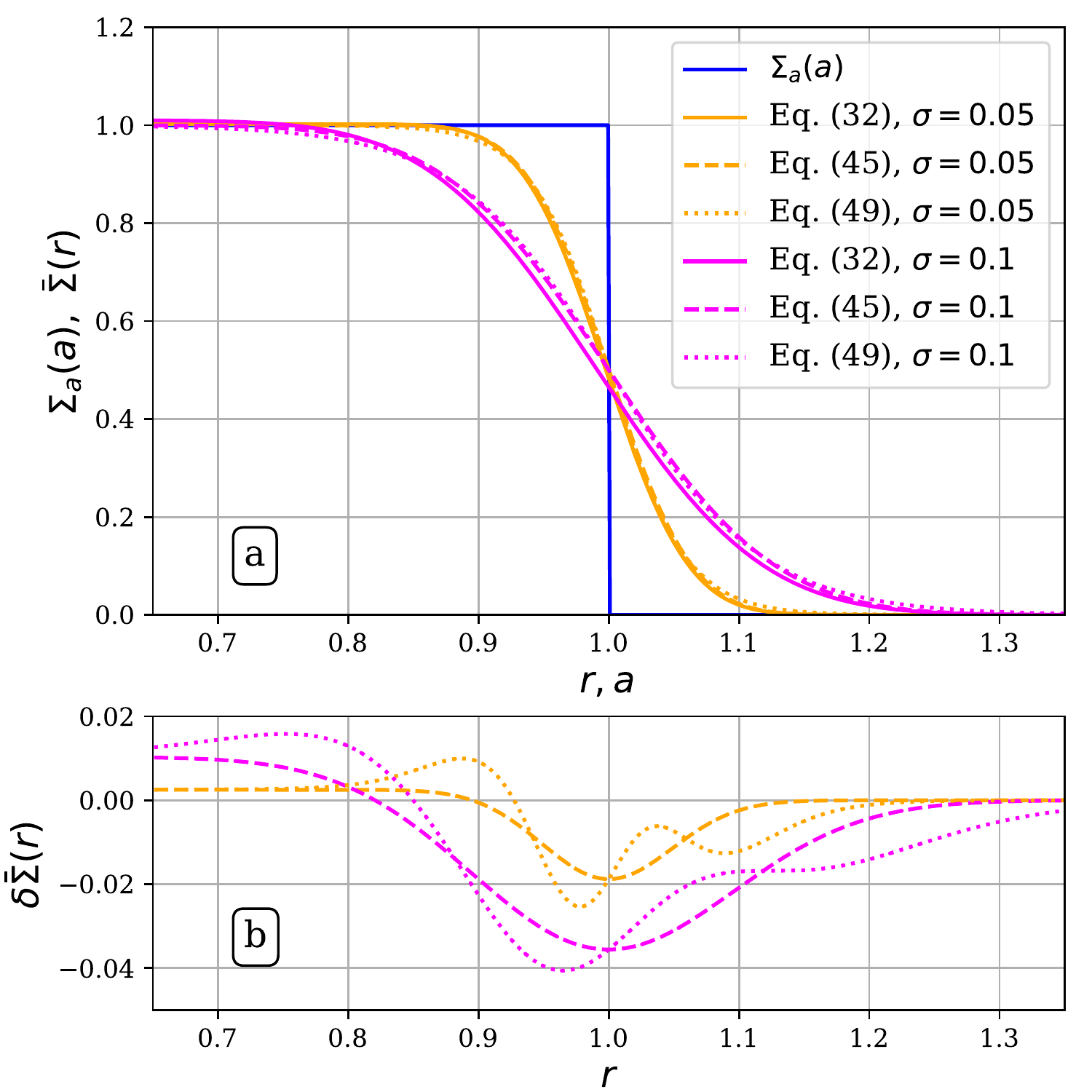}
    \caption{(a) Profiles of $\asd(r)$ near a sharp edge of $\Sigma_a(a)$ (in the form (\ref{eq:Sig-edge}) with $a_0=1$) for the Rayleigh eccentricity distribution (\ref{eq:Rayleigh}). Calculations are performed for $\sigma=0.05$ (orange) and $0.1$ (purple). Solid curves are computed using full Eq. (\ref{eq:asd-rel-Ray}), while dashed curves represent small-$e$ approximation (\ref{eq:edge-Ray}). The dotted curves (almost overlapping with the dashed ones)  show the approximation (\ref{eq:edge-M}) proposed in \citet{Marino2021}, discussed in \S\ref{sec:literature}. (b) Differences between the solid and dashed ({\it dashed}) and solid and dotted ({\it dotted}) curves from panel (a), for $\sigma=0.05$ (orange) and $0.1$ (purple).}
    \label{fig:edge-Gauss}
\end{figure}

In Figure \ref{fig:edge-Gauss} we show the ASD profile near the edge of a disc with a Rayleigh distribution of eccentricities, for two values of the eccentricity dispersion $\sigma$ (solid curves). As expected, these profiles show that a transition from $\asd\approx 0$ far outside the edge to $\asd\approx \Sigma_0$ well inside of it occurs over the characteristic distance of several $\sigma a_0$.

In the small eccentricity limit ($\sigma\ll 1$) we can derive an approximate expression for $\asd(r)$. Indeed, plugging the anzatz (\ref{eq:Sig-edge}) into the equation (\ref{eq:asd-rel-Ray_app}) and working in the vicinity of the disc edge, $|r-a_0|\ll a_0$, one finds 
\ba  
\asd(r)\approx \frac{\Sigma_0}{2}\left[1-{\rm erf} \left(\frac{r-a_0}{\sqrt{2}\sigma a_0}\right)\right].
\label{eq:edge-Ray}
\ea  
This approximation is illustrated in Figure \ref{fig:edge-Gauss}a (dashed curves), while the dashed curves in panel (b) show its deviation from the exact calculation. The discrepancy is largest in the near-edge region (similar to Figure \ref{fig:edge-e-a}) but stays below $4\%$ for $\sigma=0.1$. Thus, the approximation (\ref{eq:edge-Ray}) works reasonably well. 

Comparing Figures \ref{fig:edge-e-a} and \ref{fig:edge-Gauss}, one can see clear differences in the $\asd(r)$ profiles obtained for different particle eccentricity distributions, albeit not as dramatic as in the case of a narrow ring (see Figure \ref{sec:rings-e}). A disc with $e=e_a$ near the edge has a sharper transition of $\asd$ from $\Sigma_0$ to zero, which also occurs over a smaller radial interval than in the case of the Rayleigh $\psi_e$. While Figure \ref{fig:edge-e-a} features a very steep (formally divergent) slope of $\asd(r)$ at $r=a_0(1\pm e_a)$, the ASD profile in Figure \ref{fig:edge-Gauss} is smooth everywhere. This provides us with a way of differentiating the dynamical states of debris discs by examining their brightness profiles near the sharp edges of $\Sigma_a(a)$. 

\subsubsection{General method for retrieving $\Phi_e(r,a)$ near sharp disc edges}
\label{sec:edges-gen}


We now describe a general formalism for deriving $\Phi_e(r,a)$ from the measurement of $\asd(r)$ near a sharp edge of a disc in the form (\ref{eq:Sig-edge}). As stated earlier (and assumed in the examples shown in \S\ref{sec:edges-ea}, \ref{sec:edges-Ray}), we need to assume that $\psi_e(e,a)$ is independent of $a$ near the edge. According to equation (\ref{eq:Psi_gen}), this implies that $\Phi_e(r,a)$ depends on $r$ and $a$ only through the variable $\kappa$ given by the equation (\ref{eq:kappa}), which in particular means that  $\Phi_e(r,a)=\Phi_e(r/a)$. Then we can write the equation (\ref{eq:asd-rel-gen}) as
\ba  
\asd(r) &=& \frac{\Sigma_0}{\pi}\int_{r/2}^{a_0} a^{-1}\Phi_e\left(\frac{r}{a}\right)  \md a
\nonumber\\
&=& \frac{\Sigma_0}{\pi}\int_{1/2}^{a_0/r} x^{-1}\Phi_e\left(x^{-1}\right) \md x,
\label{eq:asd-edge}
\ea  
where in the second line we made a change of variables $a\to rx$. Differentiating equation (\ref{eq:asd-edge}) with respect to $r$ we find 
\ba  
\Phi_e(r,a)= -\frac{\pi}{\Sigma_0}\left(y\frac{\md \asd(y)}{\md y}\right)\bigg|_{y=(a_0/a)r},
\label{eq:Psi-edge}
\ea  
with $y$ being a dummy variable. Plugging this $\Phi_e(r,a)$ in the equation (\ref{eq:Abel}), one can obtain $\psi_e(e,a)=\psi_e(e,a_0)$ at the edge of the disc, completing our task. This procedure is illustrated in detail in Appendix \ref{sec:psi-edge}.

Because of the constraint (\ref{eq:symmetry}) and the relation (\ref{eq:Psi-edge}), the underlying assumption of a sharp truncation of $\Sigma_a(a)$ at some $a_0$ can be justified only if the observed $\asd(r)$ satisfies
\ba  
\left(\md \asd/\md \ln r\right)|_{r=a_0+z}=\left(\md \asd/\md \ln r\right)|_{r=a_0-z}
\label{eq:sym-edge}
\ea 
for any $z$. This condition provides a useful check of the sharp (in $a$) edge assumption: if the observationally determined $\asd(r)$ does not satisfy (\ref{eq:sym-edge}), then $\Sigma_a(a)$ cannot have a form close to (\ref{eq:Sig-edge}).

One can see that both here, as well as in the ring case considered in \S\ref{sec:rings-e}, the key to a successful retrieval of the $\psi_e$ dependence on $e$ lies in the fact that the eccentricity distribution was assumed to be independent of $a$. Thus, such procedure can inform us about the behavior of $\psi_e$ only locally, near a sharp feature (ring or edge) of $\Sigma_a(a)$. In particular, it may not work so well for discs that have been secularly evolving for much longer than the apsidal precession period at $a_0$, as in that case $e_a(a)$ would rapidly oscillate in $a$ near $a_0$ (e.g. see \citealt{Sefilian2020}).


\section{Discussion}
\label{sec:disc}


Results of our study allow one to uniquely relate the distribution of the orbital elements of debris disc particles, namely $\Sigma_a(a)$ and $\psi_e(e,a)$, to a particular observable --- azimuthally averaged surface density $\asd(r)$. Equations (\ref{eq:asd-rel-gen}) and (\ref{eq:Psi_e}) or (\ref{eq:Psi_gen}) directly link the underlying kinematic characteristics of the disc and its $\asd(r)$ profile. One can use these relations to test different assumptions about the forms of $\Sigma_a(a)$ and $\psi_e(e,a)$ by computing the corresponding $\asd(r)$ and comparing it with observations. For certain forms of particle eccentricity distribution expected in debris discs on theoretical grounds, we were able to simplify things even further, see \S\ref{sec:unique_e} for $e=e_a(a)$ and \S\ref{sec:Rayleigh} for the Rayleigh distribution.

\begin{figure}
	\includegraphics[width=\linewidth]{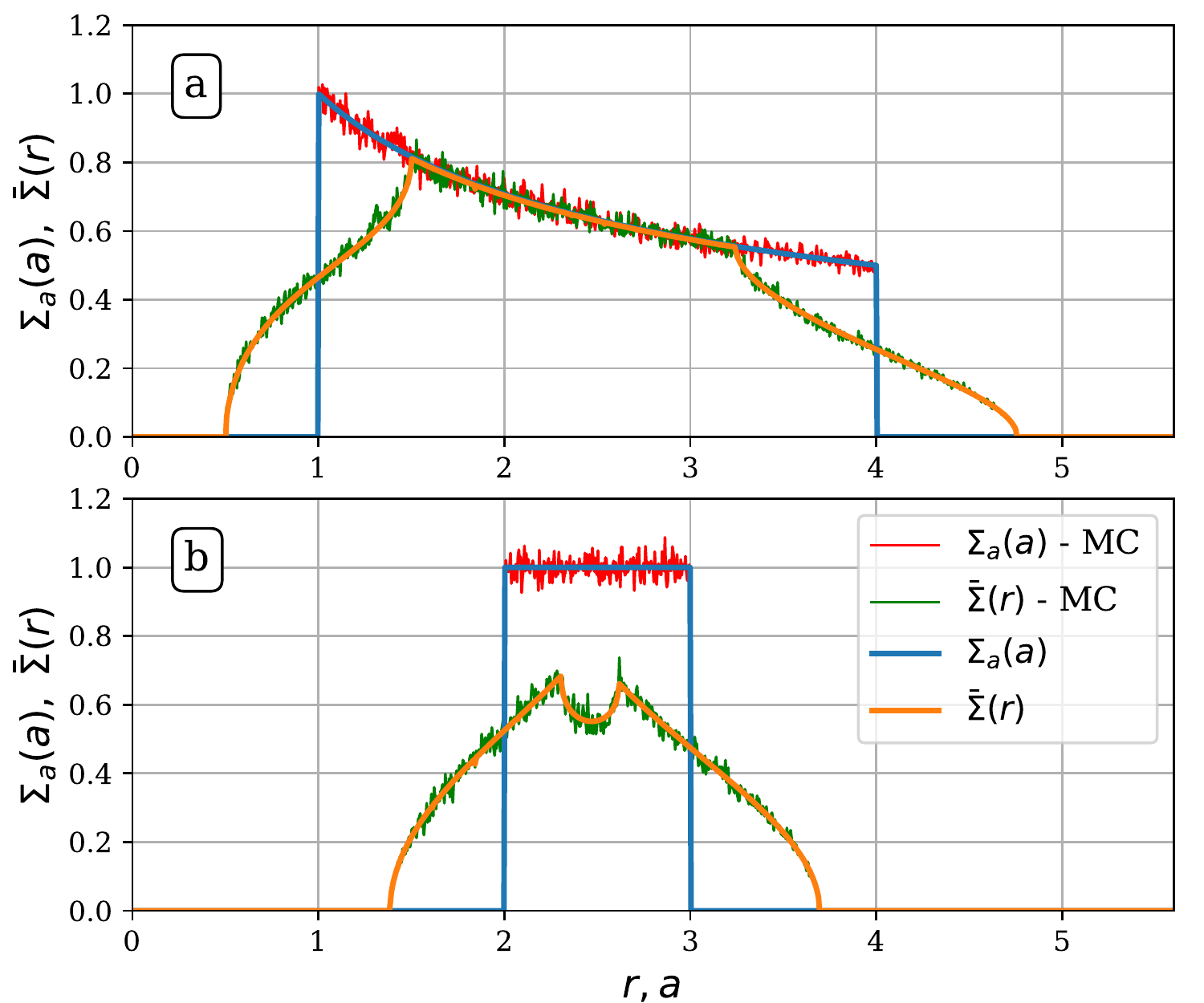}
    \caption{Comparison of $\asd(r)$ computed by our method and by Monte Carlo sampling of orbital elements, using the examples shown in Figure \ref{fig:Psi-e}b,c (current panels (a) \& (b), correspondingly). Monte Carlo sampling (thin red curves) of the underlying $\Sigma_a(a)$ (thick blue) uses semi-major axis bins of width 0.05, covering the interval $a\in (1,4)$, with a total of $6\times 10^5$ orbit samples (i.e. $10^3$ per $a$-bin). Despite the large number of samples, the resultant Monte Carlo $\asd(r)$ (thin green) is still noisy compared to our exact analytical caclulation (thick orange).}
    \label{fig:sampling}
\end{figure}

The most common method to predict $\asd(r)$ (or\footnote{Obtaining $\Sigma(r,\phi)$ also requires the knowledge of the distribution of particle apsidal angles $\varpi$.} $\Sigma(r,\phi)$) for given $\Sigma_a(a)$ and $\psi_e$ is to randomly sample these distributions in the Monte Carlo fashion assuming (as we do as well) a uniform distribution in the mean longitude of particles \citep{Lee2016,Sefilian2020,Marino2021}. The resultant instantaneous positions of the debris particles are then binned in $r$-coordinate (or in $(r,\phi)$-plane) yielding an approximation for $\asd(r)$ (or $\Sigma(r,\phi)$). This method is illustrated in Figure \ref{fig:sampling}, in which it is used for obtaining $\asd(r)$ for two disc models previously shown in Figure \ref{fig:Psi-e}b,c. For both models the underlying $\Sigma_a(a)$ (blue curve) is sampled (red curves) with $10^3$ orbits per every bin in $a$, with the bin width of $\Delta a=0.05$. Green curves show the resultant $\asd(r)$ profiles, with orange curves showing the exact calculation as in Figure \ref{fig:Psi-e}. One can see that $\asd(r)$ resulting from the Monte Carlo procedure is rather noisy despite the large number of particles ($6\times 10^5$) involved in this calculation. Moreover, even at this level of accuracy the Monte Carlo calculation takes an order of magnitude longer to compute than the exact calculation using our method, i.e. equations (\ref{eq:asd-rel-gen})-(\ref{eq:theta1}).

Another benefit of our framework is that the function $\Phi_e(r,a)$ given by equations (\ref{eq:Psi_e}), (\ref{eq:Psi_gen}) is independent of $\Sigma_a$. As a result, a single computation of $\Phi_e(r,a)$ can be reused multiple times to generate $\asd(r)$ profiles for an arbitrary number of disc models with different $\Sigma_a(a)$ via the equation (\ref{eq:asd-rel-gen}), as long as these models have the same $\psi_e(e,a)$. This makes our method very efficient when one needs to compute $\asd(r)$ for a large number of disc models with the same $\psi_e(e,a)$, e.g. when fitting observational data.

Also, one might often be interested in obtaining the value of $\asd(r)$ at a {\it single} $r$, rather than its full radial profile. To get that, Monte Carlo technique would still require sampling the orbital distribution of the debris particles over a significant portion of the disc, if not all of it. In other words, with Monte Carlo sampling the computational cost of obtaining $\asd$ at a single point is not much different from that of computing the full $\asd(r)$ profile. This is not the case for our method, which naturally allows one to compute $\asd(r)$ at a single (or a few) point(s) via the equations (\ref{eq:asd-rel-gen}), (\ref{eq:Psi_e}), (\ref{eq:Psi_gen}) with minimal numerical cost.   

As shown in \S\S\ref{sec:inverse-mass},\ref{sec:inverse-e}, the relations (\ref{eq:asd-rel-gen}), (\ref{eq:Psi_e}),  (\ref{eq:Psi_gen}) also provide a basis for {\it inverting} the problem, i.e. using disc observations to infer the distribution of debris disc particles in semi-major axis or eccentricity. Unfortunately, one cannot simultaneously obtain both $\Sigma_a(a)$ and $\psi_e(e,a)$ in a unique fashion given $\asd(r)$, and some additional considerations must be used for solving the inverse problem. Nevertheless, our results do provide explicit recipes for inferring either $\Sigma_a(a)$ (see \S\ref{sec:inverse-mass}) or, in some situations, $\psi_e(a,e)$ (see \S\ref{sec:inverse-e} and Appendices \ref{sec:psi-rings},\ref{sec:psi-edge}), given the knowledge of $\asd(r)$ (see \S\ref{sec:obs} for their practical applicability). In particular, these recipes demonstrate that the inversion techniques formulated in \S\S\ref{sec:inverse-mass},\ref{sec:inverse-e} are mathematically well-posed and can be viewed as proofs of existence of the solutions of the corresponding problems. 

Our focus on the zeroth-order azimuthal moment $\asd(r)$ of the full surface density distribution $\Sigma(r,\phi)$ in this work is not accidental. First, from the observational point of view, ASD should have a higher signal-to-noise ratio than $\Sigma(r,\phi)$, facilitating its analysis. Second, $\asd(r)$ is related to the underlying orbital properties of disc particles via the straightforward residence time argument (\S\ref{sec:unique_e}), enabling the derivation of the master equation\footnote{Although we provide its alternative derivation in Appendix \ref{sec:fluid}.} (\ref{eq:asd-rel-gen}). Third, as already noted in \S\ref{sec:master-eq}, the calculation of $\asd(r)$ is entirely decoupled from the distribution of apsidal angles of disc particles, as $\varpi$ does not enter the relation (\ref{eq:asd-rel-gen}) between $\asd(r)$ and $\Sigma_a(a)$, $\psi_e(e,a)$. 

Clearly, $\Sigma(r,\phi)$ encodes more information about kinematic properties of disc particles than $\asd(r)$ alone, simply because $\Sigma(r,\phi)$ also depends on their apsidal angles $\varpi$. With the residence time argument we cannot predict the non-axisymmetric structure of the disc, since different $\Sigma(r,\phi)$ may result in the same $\asd(r)$. For example, a ring of a kind (\ref{eq:ring}) considered in \S\ref{sec:rings-e} would appear in $(r,\phi)$-space as an elliptical wire for a tightly clustered (i.e. apsidally-aligned) $\varpi$ distribution, but as an axisymmetric ring for randomly distributed $\varpi$; in both cases $\asd(r)$ would be the same. 
Our approach turns this ambiguity into a benefit, as it effectively {\it decouples} the determination of eccentricity or mass distributions in the disc from the problem of constraining the apsidal angle distribution. In other words, our framework allows one to use observationally derived $\asd(r)$ (having improved signal-to-noise ratio compared to $\Sigma(r,\phi)$) to first determine either $\Sigma_a(a)$ or $\psi_e(e,a)$. Then this information can be used to facilitate the retrieval of the apsidal angle distribution from the {\it full} $\Sigma(r,\phi)$. Such a two-step approach may provide an efficient way of inferring all kinematic characteristics of debris discs. 

Our derivation of the master equation (\ref{eq:asd-rel-gen}) in \S\ref{sec:unique_e} relies on the assumption that debris particles move on eccentric orbits in a Newtonian potential (with randomized mean anomalies). One may wonder whether some additional physical effects may affect this assumption. Collisions between particles are certainly important for the long-term evolution of debris discs. However, as long as the characteristic time between collisions is longer than the particle orbital period, our assumption of eccentric orbits in the form (\ref{eq:ellipse}) would remain valid.

Radiative effects are also important for debris disc evolution \citep{Burns1979}. Poynting-Robertson drag is a slow process which causes only minuscule departures of particle orbits from Keplerian ellipses (especially for bigger grains) and can be neglected. Radiation pressure is unimportant for large grains such as the ones responsible for thermal emission visible in sub-mm (e.g. by {\it ALMA}), but it can play an important role in the dynamics of $\mu$m and sub-$\mu$m grains that dominate the scattered light observations. However, even for these small grains the only effect of radiation pressure is the effective reduction of the central object's mass \citep{Burns1979}, while the orbits still remain closed ellipses with all other features of Keplerian dynamics fully preserved. Since all our results, including the master equation (\ref{eq:asd-rel-gen}) are completely independent of the central mass $M_\star$, they should remain valid even for small dust particles strongly affected by radiation pressure.

Gas has been detected in a number of debris discs \citep{Hughes2018} and may affect the motion of debris particles. However, as long as the gas density is low enough and/or the emitting particle size is large enough for gas drag to not cause significant departures of their orbits from purely Keplerian, presence of gas should not affect our conclusions, similar to the Poynting-Robertson drag.


\subsection{Application to debris disc observations}
\label{sec:obs}


Application of the inversion techniques described in \S\S\ref{sec:inverse-mass},\ref{sec:inverse-e} to observations of debris discs can shed light on their architecture at formation, as well as on the dynamical processes operating in them. In particular, knowledge of $\Sigma_a(a)$ for discs undergoing long-term secular evolution translates into the knowledge of initial mass distribution in the disc, as particle semi-major axes do not change in the course of secular evolution. Thus, if the disc started off dynamically cold, with $e\to 0$ everywhere, and then attained its current (non-zero) eccentricity distribution as a result of secular evolution (e.g. due to a planetary or stellar perturber), then the initial $\Sigma(r,\phi)|_{t=0}=\asd(r)|_{t=0}$ of the disc and its current $\Sigma_a(a)$ must be related as $\Sigma(r,\phi)|_{t=0}=\Sigma_a(r)$. 

Constraints on the shape of $\psi_e(e,a)$, even if obtained only locally, like in \S\ref{sec:inverse-e}, can shed light on the processes that govern eccentricities of the debris disc particles. For example, a truly smooth and extended (in $e$) $\psi_e$ close in form to the Rayleigh distribution (\S\ref{sec:Rayleigh}) would support the idea that particle eccentricities are excited by disc self-stirring \citep{KenW2010}.  

On the other hand, if $\psi_e(e,a)$ is found to be clustered around a particular value of $e$, i.e. being close to $e=e_a(a)$ anzatz (\S\ref{sec:unique_e}), this might indicate the prevalence of secular processes in determining eccentricity dynamics. In the case of a spatially extended disc this would also suggest that the disc is "secularly young", i.e. its age is not very large compared to the local apsidal precession period, as otherwise $e_a(a)$ would rapidly oscillate in $a$ (as we alluded to in \S\ref{sec:edges-gen}), potentially masquerading as an extended and continuous (in $e$) $\psi_e$.

It should be noted that some of the inversion recipes proposed in \S\S\ref{sec:inverse-mass},\ref{sec:inverse-e} may have somewhat limited practical utility when applied to real observational data, simply because the latter is intrinsically noisy. For example, re-construction of $\Phi_e(r,a_0)$ near a sharp edge of a disc at $a=a_0$ via the equation (\ref{eq:Psi-edge}) involves a differentiation of (noisy) $\asd(r)$ derived from observations, which is going to amplify the noise. Subsequent retrieval of $\psi_e(a_r,e)$ from this $\Phi_e(r,a_0)$ using equation (\ref{eq:Abel}) involves yet another differentiation, exacerbating the noise issue even further. Thus, one should be cautious when directly applying equations (\ref{eq:volt})-(\ref{eq:Abel}), (\ref{eq:Psi-edge}) to real data, which may need to be smoothed prior to using this technique. 

In many cases a forward modeling approach, i.e. varying the parameters of a model for $\Sigma_a(a)$ or $\psi_e(a,e)$ until $\asd(r)$ predicted by the equation (\ref{eq:asd-rel-gen}) best matches the data, would be more appropriate. Our analytical framework provides a very efficient tool for such forward modeling, see the earlier discussion of Figure \ref{fig:sampling}. 

To obtain a reliable $\asd(r)$ from observations one needs to have an accurate mapping of the surface density in $(r,\phi)$ coordinates in the plane of the disc in the first place. This requires de-projecting the observed brightness distribution in the plane of the sky, accounting for the overall disc inclination, and converting brightness distribution to $\Sigma(r,\phi)$. The latter should be relatively straightforward for systems imaged in sub-mm with e.g. {\it ALMA}, when one expects particle brightness to decay as $r^{-1/2}$ in the optically thin, Rayleigh-Jeans regime. Things are more complicated for systems imaged in the optical and near-IR through scattered light\footnote{It should be remembered that thermal sub-mm emission and scattered light observations are probing very different dust populations, which may have rather different kinematic properties.}, as in this case one also need to account for the scattering phase function. 

Disc warping might complicate de-projection, especially for discs which are close to edge-on. If a disc has a substantial spread of particle inclinations around its midplane, this would be another complication (especially for close to edge-on discs, see \citealt{Han2022}), although less severe for almost face-on systems for which inclination enters the determination of $\Sigma(r,\phi)$ only as a second order effect compared to eccentricity \citep{Marino2021}. And having gone through these procedures, one would obtain $\Sigma(r,\phi)$ for the luminous dust particles only --- the actual mass surface density distributed in objects of all sizes may be quite different from this $\Sigma(r,\phi)$. 

Despite these complications, ASD profiles have been derived and analyzed for a number of debris discs (e.g. \citealt{Marino2021}), enabling application of our technique to real observations.


\subsection{Application to other astrophysical discs/rings}
\label{sec:other}


Throughout this work we focused on modeling debris discs. However, our framework can be applied to other types of discs/rings as well. For example, particles in planetary rings with low optical depth may experience collisions rarely enough for the assumptions of their Keplerian motion and randomized mean longitudes to be valid. In such tenuous rings our results would apply for a properly measured ASD.  

Large dust grains in the outer parts of protoplanetary discs are weakly affected by gas drag (if their Stokes number is much larger than unity) and move on essentially Keplerian orbits. Thus, our methods would apply for such discs, even though particle motion in them is unlikely to strongly deviate from the gas motion. 

Another example of relevant nearly-Keplerian systems is provided by the nuclear stellar discs orbiting supermassive black holes in centers of galaxies, with stars replacing debris particles. In particular, the eccentric stellar disc in the center of M31 \citep{Tremaine1995} has been a subject of numerous observations \citep{Lauer1993,Lauer1998} and extensive modeling \citep{Salow2001,Salow2004,Peiris2003,Brown2013}. Our framework can be directly applied to modeling this system. This disc has a particular benefit of also having a resolved kinematic information available to us \citep{Bacon2001,Bender2005}, providing an additional constraint on the semi-major axis and eccentricity distributions $\Sigma_a(a)$ and $\psi_e(a,e)$. 

Finally, we note that the residence time argument can be applied even to non-Keplerian discs, as long as their potential can be approximated as axisymmetric in the disc plane. For example, it can be employed to provide a relation similar to the equation (\ref{eq:asd-rel-gen}) between the stellar energy and angular momentum distributions in disc galaxies and their (azimuthally-averaged) surface density. It should be remembered, however, that the form of the underlying potential is not known a priori in this case (but kinematic data might help in constraining it).


\subsection{Comparison with the existing studies}
\label{sec:literature}


A number of past studies computed $\Sigma(r,\phi)$ by randomly sampling some prescribed distributions of $a$, $e$, and $\varpi$ \citep{Wyatt2005,Lee2016}, as described in the discussion of Figure \ref{fig:sampling}. In some cases the radial profiles of ASD were also computed using Monte Carlo sampling \citep{Yel2018,Sefilian2020}.

\citet{Marino2021} used Monte Carlo sampling to study the characteristics of radial density profiles of debris discs with the Rayleigh distribution of eccentricities and random $\varpi$. Results of this exercise have been used to constrain eccentricity dispersion $\sigma$ in several debris discs observed with {\it ALMA}. A disc with a uniformly distributed $\varpi$ is axisymmetric and its $\Sigma(r,\varphi)$ naturally coincides with $\asd(r)$. For that reason we can directly apply our results to provide some insights into the calculations of \citet{Marino2021}.

In particular, \citet{Marino2021} explored the ASD profile near a sharp edge of a disc, where $\Sigma_a(a)$ is of the form (\ref{eq:Sig-edge}). They came up with the following fitting function to describe the behavior of $\asd(r)$ near the (outer) edge:
\ba  
\asd(r)\approx \frac{\Sigma_0}{2}\left[1+{\rm tanh} \left(\frac{a_0-r}{l_\mathrm{out}}\right)\right],
\label{eq:edge-M}
\ea  
with $l_\mathrm{out}\approx e_\mathrm{rms} a_0/1.2$, where $e_\mathrm{rms}=\sqrt{2}\sigma$ in our notation. 

In Figure \ref{fig:edge-Gauss} we compare the performance of this fitting formula (dotted curves) with the exact calculation using equation (\ref{eq:asd-rel-Ray}), as well as with our low-$\sigma$ approximation near the edge (\ref{eq:edge-Ray}). Interestingly, the fitting formula (\ref{eq:edge-M}) works reasonably well, with deviations from the exact $\asd(r)$ being comparable to the deviations of our (physically motivated) low-$\sigma$ approximation (\ref{eq:edge-Ray}), see Figure \ref{fig:edge-Gauss}b. Thus, the fitting formula (\ref{eq:edge-M}) of \citet{Marino2021} can be used on par with the equation (\ref{eq:edge-Ray}) to characterize ASD profile near the edge of a debris disc.


\section{Summary}
\label{sec:Summary}


In this work we developed a (semi-)analytical formalism allowing one to relate the (arbitrary) semi-major axis and eccentricity distributions of particles in debris discs to their azimuthally-averaged surface density, which can be inferred from the brightness distribution of the disc. Our method can be used for fast and efficient forward-modeling of debris disc observations, allowing one to compute multiple surface density profiles for a given eccentricity distribution. Alternatively, in certain situations our framework can be used for retrieving orbital parameter distributions of the disc particles directly from observations (\S\S\ref{sec:inverse-mass},\ref{sec:inverse-e}). Explicit formulae have been derived for a number of eccentricity distributions $\psi_e$, including a prescribed eccentricity profile $e_a(a)$ and the Rayleigh distribution. This framework does not require the knowledge of the distribution of apsidal angles of disc particles, simplifying the inference of orbital parameters. It can also be applied to understanding the architecture and kinematic properties of other disc-like systems, e.g. nuclear stellar discs in centers of galaxies harboring supermassive black holes. Future work will aim to use the full $\Sigma(r,\phi)$ information to further explore the underlying disc kinematics, beyond the level presented in this study.


\section*{Acknowledgements}

R.R.R. acknowledges financial support through the STFC grant ST/T00049X/1 and Ambrose Monell Foundation. 

\section*{Data availability}

Data produced in this paper is available from the author upon reasonable request.



\bibliographystyle{mnras}
\bibliography{Bibliography}

\begin{thebibliography}{}
\makeatletter
\relax
\def\mn@urlcharsother{\let\do\@makeother \do\$\do\&\do\#\do\^\do\_\do\%\do\~}
\def\mn@doi{\begingroup\mn@urlcharsother \@ifnextchar [ {\mn@doi@}
  {\mn@doi@[]}}
\def\mn@doi@[#1]#2{\def\@tempa{#1}\ifx\@tempa\@empty \href
  {http://dx.doi.org/#2} {doi:#2}\else \href {http://dx.doi.org/#2} {#1}\fi
  \endgroup}
\def\mn@eprint#1#2{\mn@eprint@#1:#2::\@nil}
\def\mn@eprint@arXiv#1{\href {http://arxiv.org/abs/#1} {{\tt arXiv:#1}}}
\def\mn@eprint@dblp#1{\href {http://dblp.uni-trier.de/rec/bibtex/#1.xml}
  {dblp:#1}}
\def\mn@eprint@#1:#2:#3:#4\@nil{\def\@tempa {#1}\def\@tempb {#2}\def\@tempc
  {#3}\ifx \@tempc \@empty \let \@tempc \@tempb \let \@tempb \@tempa \fi \ifx
  \@tempb \@empty \def\@tempb {arXiv}\fi \@ifundefined
  {mn@eprint@\@tempb}{\@tempb:\@tempc}{\expandafter \expandafter \csname
  mn@eprint@\@tempb\endcsname \expandafter{\@tempc}}}

\bibitem[\protect\citeauthoryear{{Bacon}, {Emsellem}, {Combes}, {Copin},
  {Monnet}  \& {Martin}}{{Bacon} et~al.}{2001}]{Bacon2001}
{Bacon} R.,  {Emsellem} E.,  {Combes} F.,  {Copin} Y.,  {Monnet} G.,   {Martin}
  P.,  2001, \mn@doi [\aap] {10.1051/0004-6361:20010317}, \href
  {https://ui.adsabs.harvard.edu/abs/2001A&A...371..409B} {371, 409}

\bibitem[\protect\citeauthoryear{{Bender} et~al.,}{{Bender}
  et~al.}{2005}]{Bender2005}
{Bender} R.,  et~al., 2005, \mn@doi [\apj] {10.1086/432434}, \href
  {https://ui.adsabs.harvard.edu/abs/2005ApJ...631..280B} {631, 280}

\bibitem[\protect\citeauthoryear{{Binney} \& {Tremaine}}{{Binney} \&
  {Tremaine}}{2008}]{Binney2008}
{Binney} J.,  {Tremaine} S.,  2008, {Galactic Dynamics: Second Edition}.
Princeton University Press

\bibitem[\protect\citeauthoryear{{Brown} \& {Magorrian}}{{Brown} \&
  {Magorrian}}{2013}]{Brown2013}
{Brown} C.~K.,  {Magorrian} J.,  2013, \mn@doi [\mnras] {10.1093/mnras/stt104},
  \href {https://ui.adsabs.harvard.edu/abs/2013MNRAS.431...80B} {431, 80}

\bibitem[\protect\citeauthoryear{{Burns}, {Lamy}  \& {Soter}}{{Burns}
  et~al.}{1979}]{Burns1979}
{Burns} J.~A.,  {Lamy} P.~L.,   {Soter} S.,  1979, \mn@doi [\icarus]
  {10.1016/0019-1035(79)90050-2}, \href
  {https://ui.adsabs.harvard.edu/abs/1979Icar...40....1B} {40, 1}

\bibitem[\protect\citeauthoryear{Gradshteyn \& Ryzhik}{Gradshteyn \&
  Ryzhik}{2014}]{Gradshteyn2014}
Gradshteyn I.~S.,  Ryzhik I.~M.,  2014, Table of integrals, series, and
  products.
Academic press

\bibitem[\protect\citeauthoryear{{Han}, {Wyatt}  \& {Matr{\`a}}}{{Han}
  et~al.}{2022}]{Han2022}
{Han} Y.,  {Wyatt} M.~C.,   {Matr{\`a}} L.,  2022, \mn@doi [\mnras]
  {10.1093/mnras/stac373}, \href
  {https://ui.adsabs.harvard.edu/abs/2022MNRAS.511.4921H} {511, 4921}

\bibitem[\protect\citeauthoryear{{Hughes}, {Duch{\^e}ne}  \&
  {Matthews}}{{Hughes} et~al.}{2018}]{Hughes2018}
{Hughes} A.~M.,  {Duch{\^e}ne} G.,   {Matthews} B.~C.,  2018, \mn@doi [\araa]
  {10.1146/annurev-astro-081817-052035}, \href
  {https://ui.adsabs.harvard.edu/abs/2018ARA&A..56..541H} {56, 541}

\bibitem[\protect\citeauthoryear{{Ida} \& {Makino}}{{Ida} \&
  {Makino}}{1992}]{Ida1992}
{Ida} S.,  {Makino} J.,  1992, \mn@doi [\icarus]
  {10.1016/0019-1035(92)90008-U}, \href
  {https://ui.adsabs.harvard.edu/abs/1992Icar...96..107I} {96, 107}

\bibitem[\protect\citeauthoryear{{Ida}, {Kokubo}  \& {Makino}}{{Ida}
  et~al.}{1993}]{IdaKM1993}
{Ida} S.,  {Kokubo} E.,   {Makino} J.,  1993, \mn@doi [\mnras]
  {10.1093/mnras/263.4.875}, \href
  {https://ui.adsabs.harvard.edu/abs/1993MNRAS.263..875I} {263, 875}

\bibitem[\protect\citeauthoryear{{Kennedy}}{{Kennedy}}{2020}]{Kennedy2020}
{Kennedy} G.~M.,  2020, \mn@doi [Royal Society Open Science]
  {10.1098/rsos.200063}, \href
  {https://ui.adsabs.harvard.edu/abs/2020RSOS....700063K} {7, 200063}

\bibitem[\protect\citeauthoryear{{Kennedy} \& {Wyatt}}{{Kennedy} \&
  {Wyatt}}{2010}]{KenW2010}
{Kennedy} G.~M.,  {Wyatt} M.~C.,  2010, \mn@doi [\mnras]
  {10.1111/j.1365-2966.2010.16528.x}, \href
  {https://ui.adsabs.harvard.edu/abs/2010MNRAS.405.1253K} {405, 1253}

\bibitem[\protect\citeauthoryear{{Lauer} et~al.,}{{Lauer}
  et~al.}{1993}]{Lauer1993}
{Lauer} T.~R.,  et~al., 1993, \mn@doi [\aj] {10.1086/116737}, \href
  {https://ui.adsabs.harvard.edu/abs/1993AJ....106.1436L} {106, 1436}

\bibitem[\protect\citeauthoryear{{Lauer}, {Faber}, {Ajhar}, {Grillmair}  \&
  {Scowen}}{{Lauer} et~al.}{1998}]{Lauer1998}
{Lauer} T.~R.,  {Faber} S.~M.,  {Ajhar} E.~A.,  {Grillmair} C.~J.,   {Scowen}
  P.~A.,  1998, \mn@doi [\aj] {10.1086/300617}, \href
  {https://ui.adsabs.harvard.edu/abs/1998AJ....116.2263L} {116, 2263}

\bibitem[\protect\citeauthoryear{{Lee} \& {Chiang}}{{Lee} \&
  {Chiang}}{2016}]{Lee2016}
{Lee} E.~J.,  {Chiang} E.,  2016, \mn@doi [\apj] {10.3847/0004-637X/827/2/125},
  \href {https://ui.adsabs.harvard.edu/abs/2016ApJ...827..125L} {827, 125}

\bibitem[\protect\citeauthoryear{{Marino}}{{Marino}}{2021}]{Marino2021}
{Marino} S.,  2021, \mn@doi [\mnras] {10.1093/mnras/stab771}, \href
  {https://ui.adsabs.harvard.edu/abs/2021MNRAS.503.5100M} {503, 5100}

\bibitem[\protect\citeauthoryear{{Murray} \& {Dermott}}{{Murray} \&
  {Dermott}}{1999}]{Murray1999}
{Murray} C.~D.,  {Dermott} S.~F.,  1999, {Solar system dynamics}

\bibitem[\protect\citeauthoryear{{Pearce} \& {Wyatt}}{{Pearce} \&
  {Wyatt}}{2014}]{Pearce2014}
{Pearce} T.~D.,  {Wyatt} M.~C.,  2014, \mn@doi [\mnras]
  {10.1093/mnras/stu1302}, \href
  {https://ui.adsabs.harvard.edu/abs/2014MNRAS.443.2541P} {443, 2541}

\bibitem[\protect\citeauthoryear{{Peiris} \& {Tremaine}}{{Peiris} \&
  {Tremaine}}{2003}]{Peiris2003}
{Peiris} H.~V.,  {Tremaine} S.,  2003, \mn@doi [\apj] {10.1086/378638}, \href
  {https://ui.adsabs.harvard.edu/abs/2003ApJ...599..237P} {599, 237}

\bibitem[\protect\citeauthoryear{{Press}, {Teukolsky}, {Vetterling}  \&
  {Flannery}}{{Press} et~al.}{2002}]{NR2002}
{Press} W.~H.,  {Teukolsky} S.~A.,  {Vetterling} W.~T.,   {Flannery} B.~P.,
  2002, {Numerical recipes in C++ : the art of scientific computing}

\bibitem[\protect\citeauthoryear{{Salow} \& {Statler}}{{Salow} \&
  {Statler}}{2001}]{Salow2001}
{Salow} R.~M.,  {Statler} T.~S.,  2001, \mn@doi [\apjl] {10.1086/319834}, \href
  {https://ui.adsabs.harvard.edu/abs/2001ApJ...551L..49S} {551, L49}

\bibitem[\protect\citeauthoryear{{Salow} \& {Statler}}{{Salow} \&
  {Statler}}{2004}]{Salow2004}
{Salow} R.~M.,  {Statler} T.~S.,  2004, \mn@doi [\apj] {10.1086/422163}, \href
  {https://ui.adsabs.harvard.edu/abs/2004ApJ...611..245S} {611, 245}

\bibitem[\protect\citeauthoryear{{Sefilian}, {Rafikov}  \& {Wyatt}}{{Sefilian}
  et~al.}{2021}]{Sefilian2020}
{Sefilian} A.~A.,  {Rafikov} R.~R.,   {Wyatt} M.~C.,  2021, \mn@doi [\apj]
  {10.3847/1538-4357/abda46}, \href
  {https://ui.adsabs.harvard.edu/abs/2021ApJ...910...13S} {910, 13}

\bibitem[\protect\citeauthoryear{{Statler}}{{Statler}}{2001}]{Statler01}
{Statler} T.~S.,  2001, \mn@doi [\aj] {10.1086/323713}, \href
  {http://adsabs.harvard.edu/abs/2001AJ....122.2257S} {122, 2257}

\bibitem[\protect\citeauthoryear{{Tremaine}}{{Tremaine}}{1995}]{Tremaine1995}
{Tremaine} S.,  1995, \mn@doi [\aj] {10.1086/117548}, \href
  {https://ui.adsabs.harvard.edu/abs/1995AJ....110..628T} {110, 628}

\bibitem[\protect\citeauthoryear{{Wyatt}}{{Wyatt}}{2005}]{Wyatt2005}
{Wyatt} M.~C.,  2005, \mn@doi [\aap] {10.1051/0004-6361:20053391}, \href
  {https://ui.adsabs.harvard.edu/abs/2005A&A...440..937W} {440, 937}

\bibitem[\protect\citeauthoryear{{Wyatt}}{{Wyatt}}{2008}]{Wyatt2008}
{Wyatt} M.~C.,  2008, \mn@doi [\araa] {10.1146/annurev.astro.45.051806.110525},
  \href {https://ui.adsabs.harvard.edu/abs/2008ARA&A..46..339W} {46, 339}

\bibitem[\protect\citeauthoryear{{Yelverton} \& {Kennedy}}{{Yelverton} \&
  {Kennedy}}{2018}]{Yel2018}
{Yelverton} B.,  {Kennedy} G.~M.,  2018, \mn@doi [\mnras]
  {10.1093/mnras/sty1678}, \href
  {https://ui.adsabs.harvard.edu/abs/2018MNRAS.479.2673Y} {479, 2673}

\makeatother
\end{thebibliography}


\appendix


\section{Alternative derivation of the master equation}
\label{sec:fluid}


\citet{Statler01} provides the following expression for the full two-dimensional distribution of the surface density for a fluid disc with $e=e_a(a)$ and $\varpi=\varpi_a(a)$:
\begin{align}  
\Sigma(r,\phi) &= \Sigma_a(a)(1-e_a^2)^{1/2}D^{-1},~~~
\mbox{with}
\label{eq:Statler}
\\
D &= 1-e_a^2-e_a^\prime\left[2e_aa+r\cos(\phi-\varpi_a)\right]
\nonumber\\
& -  re_a\varpi_a^\prime\sin(\phi-\varpi_a),
\nonumber
\end{align} 
where we use $\Sigma_a(a)$ instead of $\mu(a)$ and prime denotes differentiation with respect to $a$. In this formula $a$ is a solution of the equation (\ref{eq:ellipse}) for given $r$, $\phi$, $e_a(a)$, and $\varpi_a(a)$. 

For fluid discs with no crossing of the fluid element orbits (which would otherwise result in shocks and violation of the purely Keplerian description of the orbits) this solution is unique and one must have $D>0$ keeping $\Sigma(r,\phi)$ finite. This does not have to be the case for particulate (e.g. debris) discs, in which orbit crossings are possible, potentially resulting in local singularities (caustics) of $\Sigma(r,\phi)$. In such discs the equation (\ref{eq:Statler}) gets modified: $D$ gets replaced with $|D|$, and the right hand side may involve a summation over (possibly multiple) values of $a$ satisfying the equation (\ref{eq:ellipse}) for a given $r$ and $\phi$. But, for simplicity, let us consider $\Sigma(r,\phi)$ in the form (\ref{eq:Statler}) for now. 

To get $\asd(r)$ we need to integrate $\Sigma(r,\phi)$ over $\phi$ in equation (\ref{eq:asd}), which can be turned into integration over $a$, since at a fixed $r$ a value of $\phi$ corresponds to some $a$ (modulo the insignificant $\varpi$ ambiguity) according to the equation (\ref{eq:ellipse}). In particular,
it is easy to show that 
\ba  
\md\phi=-\md a \frac{\Theta(e_a-\kappa)}{r e_a \sin(\phi-\varpi_a)}D, 
\label{eq:a-phi}
\ea  
where the introduction of the $\Theta$-function ensures that only the orbits crossing the radius $r$ contribute to $\asd$. We can also use equation (\ref{eq:ellipse}) to express
\ba  
re_a\sin(\phi-\varpi_a)=\pm a\sqrt{1-e_a^2}\sqrt{e_a^2-\kappa^2},
\label{eq:sin}
\ea  
where $\kappa$ is given by the equation (\ref{eq:kappa}). Plugging these results into the equation (\ref{eq:asd}) with $\Sigma(r,\phi)$ in the form (\ref{eq:Statler}), and multiplying $\md a$ by 2 to account for two crossings of a circle of radius $r$ by an eccentric orbit, we finally obtain (choosing the same integration limits as in \S\ref{sec:master-eq}) 
\ba  
\asd(r)=\pi^{-1}\int_{r/2}^\infty
\frac{\Sigma_a(a)}{a}\frac{\Theta(e_a-\kappa)}{\sqrt{e_a^2-\kappa^2}}\md a,
\label{eq:alt}
\ea  
in agreement with equations  (\ref{eq:asd-rel-gen})-(\ref{eq:theta1}).


\section{Power law discs}
\label{sec:PL}


\begin{figure}
	\includegraphics[width=\linewidth]{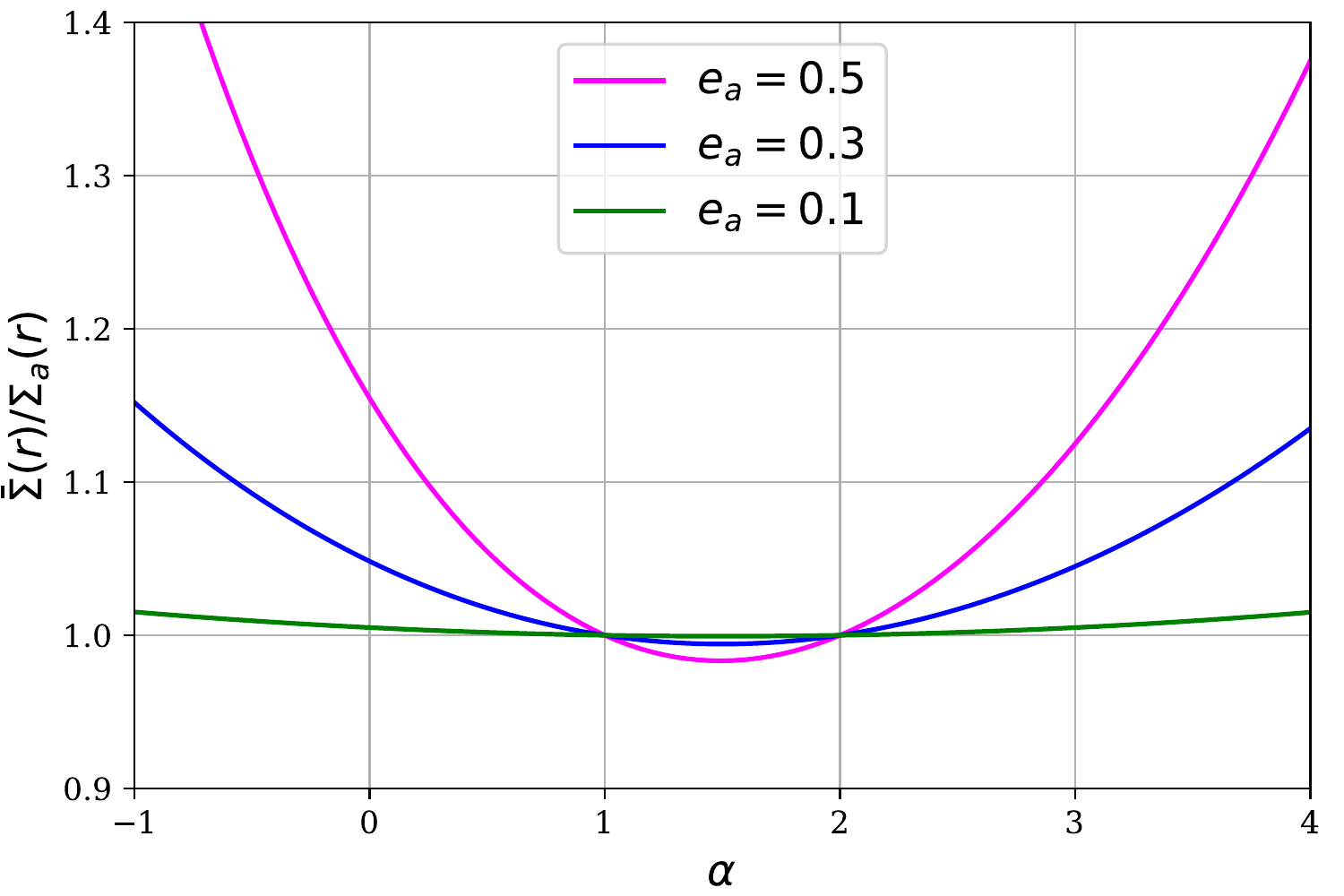}
    \caption{Ratio $\asd(r)/\Sigma_a(r)$ for a constant-$e_a$ power law disc (i.e. for $\beta=0$), shown for several values of $e_a$ (see legend) as a function of the surface density power-law index $\alpha$. }
    \label{fig:bet0}
\end{figure}

\begin{figure*}
	\includegraphics[width=\linewidth]{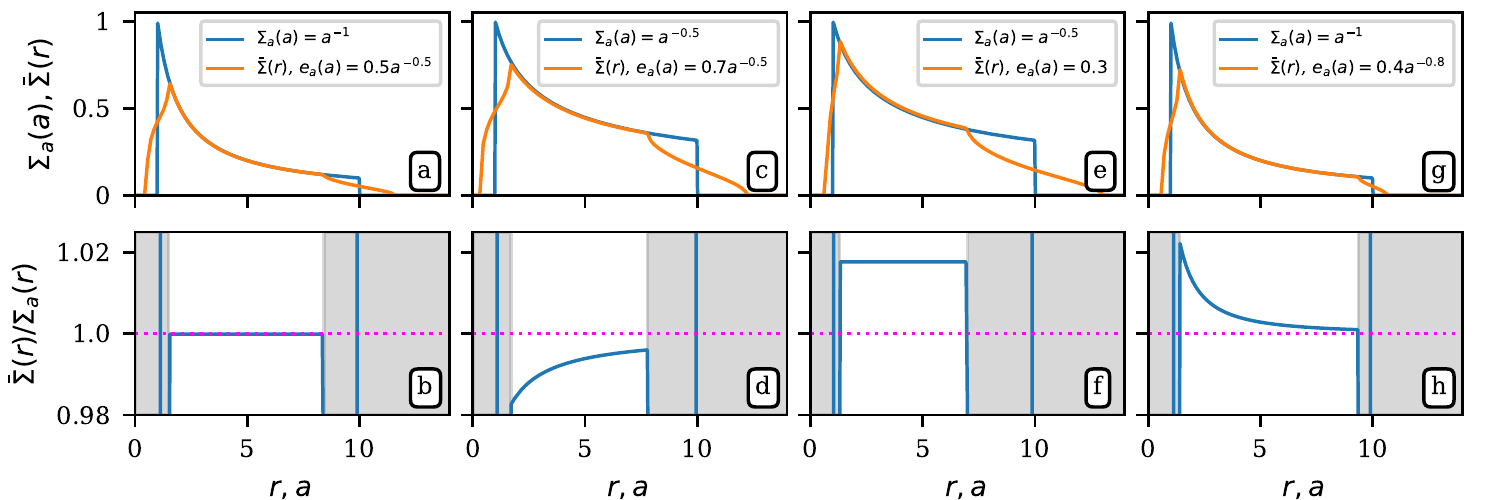}
    \caption{Behavior of (top) $\asd(r)$, $\Sigma_a(a)$ and (bottom) their ratio for different power law disc models in the form (\ref{eq:Sigma_model-pl}) and (\ref{eq:e_pl}): (a,b) ${\cal M}_\Sigma=(1,1,1,10), {\cal M}_e=(0.5,0.5,1)$, (c,d) ${\cal M}_\Sigma=(0.5,1,1,10), {\cal M}_e=(0.5,0.7,1)$, (e,f) ${\cal M}_\Sigma=(0.5,1,1,10), {\cal M}_e=(0,0.3,1)$, (g,h) ${\cal M}_\Sigma=(1,1,1,10), {\cal M}_e=(0.8,0.4,1)$, also indicated in the legends of the panels. Shaded regions in the bottom plots are affected by the edge effects and should be ignored.}
    \label{fig:PLdiscs}
\end{figure*}

Here we provide some details on the behavior of $\asd(r)$ for power-law discs with $\Sigma_a(a)$ in the form (\ref{eq:Sigma_model-pl}) and $e=e_a(a)$ in the form (\ref{eq:e_pl}). Introducing a variable $x=a/r$ we rewrite equations (\ref{eq:asd-rel-gen})-(\ref{eq:Psi_e}) as
\ba  
\asd(r) = \frac{\Sigma_a(r)}{\pi}\int_{x_1}^{x_2} \frac{x^{-1-\alpha}\md x}{\sqrt{e_a^2(r)x^{-2\beta}-(1-x^{-1})^2}},
\label{eq:asd-PL}
\ea  
where $x_{1,2}$ are the two solutions of the equation
\ba  
x^{\beta-1}|x-1|=e_a(r),
\label{eq:x_sol}
\ea  
provided that they exist. 

There are several cases in which we can obtain a more explicit relation between $\Sigma_a(a)$ and $\asd(r)$ for power law discs. This relation has a general form 
\ba  
\asd(r)=\Sigma_a(r)\varphi(r)F(r,\alpha),
\label{eq:rel-PL}
\ea  
with 
\ba  
F(r,\zeta)=\pi^{-1}\int_0^\pi\left[1+\lambda(r)\cos t\right]^{-\zeta}dt,
\label{eq:F}
\ea  
where $\varphi(r)$ and $\lambda(r)$ are certain functions determined by the value of $\beta$.

In particular, for $\beta=0$ 
\ba
\varphi(r)=\left(1-e_a^2\right)^{\alpha-1/2},~~~
\lambda(r)=e_a;
\label{eq:b0}
\ea  
are independent of $r$ since $e_a$ is a constant; this implies that $F(r,\alpha)$ is a constant as well. As a result, $\asd(r)/\Sigma_a(r)$ is independent of $r$ and is a function of $e_a$ and $\alpha$ only. In Figure \ref{fig:bet0} we show this ratio $\asd(r)/\Sigma_a(r)=\varphi(r)F(r,\alpha)$ for $\beta=0$ and several values of $e_a$, as a function of $\alpha$. For small enough $e_a\lesssim 0.1$ this ratio does not deviate from unity by more than several per cent. 

For $\beta=1/2$ one finds
\ba
\varphi(r)=\left[1+\frac{e_a^2(r)}{2}\right]^{-\alpha},~~~
\lambda(r)=e_a(r)\frac{\sqrt{1+e_a^2(r)/4}}{1+e_a^2(r)/2}.
\label{eq:b0.5}
\ea  
And for $\beta=1$ it is easy to show that $\varphi(r)=1$, $\lambda(r)=e_a(r)$. It may be possible to obtain analytical results for $\asd/\Sigma_a$ for other values of $\beta$, but we will not pursue this here. 

We illustrate these results by showing several examples of $\Sigma_a$, $\asd$ and their ratios in four power law discs in Figure \ref{fig:PLdiscs}. One can see that in general the ratio $\asd(r)/\Sigma_a(r)$ in the power law region of the disc (unshaded) is a function of $r$, and can be both larger (e.g. panels (e)-(h)) and smaller (panels (c)-(d)) than unity. Also, for some combinations of $\alpha$, $\beta$ the ratio $\asd(r)/\Sigma_a(r)$ is constant. This is certainly true for discs with $\beta=0$, i.e. having constant eccentricity, for arbitrary $\alpha$, see Equation (\ref{eq:b0}). Figure \ref{fig:PLdiscs}e,f shows an example of such a disc with $\asd(r)/\Sigma_a(r)\approx 1.018$, in agreement with Figure \ref{fig:bet0}. Also, using the fact that $F(r,0)=1$, $F(r,1)=\left[1-\lambda^2(r)\right]^{-1/2}$,
$F(r,2)=\left[1-\lambda^2(r)\right]^{-3/2}$ \citep{Gradshteyn2014}, one can easily show that $\asd(r)/\Sigma_a(r)=1$ for $(\alpha,\beta)=(0,1)$, $(1,1/2)$ (model shown in Figure (\ref{fig:PLdiscs})a,b), $(1,0)$, $(2,0)$.

Note that the model of a disc with the sharp outer edge and constant $e$ everywhere shown in Figure \ref{fig:edge-e-a} is a power law model with ${\cal M}_\Sigma=(0,1,0,1)$ and  ${\cal M}_e=(0,e_a,1)$. According to equations (\ref{eq:rel-PL})-(\ref{eq:b0}), in such a disc $F=1$ and $\asd(r)/\Sigma_a(r)=(1-e_a^2)^{-1/2}>1$ is a constant. In particular, $\asd(r)/\Sigma_a(r)\approx 1.005$ for $e_a=0.1$ and $\asd(r)/\Sigma_a(r)\approx 1.00125$ for $e_a=0.05$. This is in perfect agreement with the vertical offsets from unity of the difference curves for $r<r_p=1-e_a$ in Figure \ref{fig:edge-e-a}b.


\section{Reconstruction of $\psi_e$ for debris rings}
\label{sec:psi-rings}


\begin{figure*}
	\includegraphics[width=\linewidth]{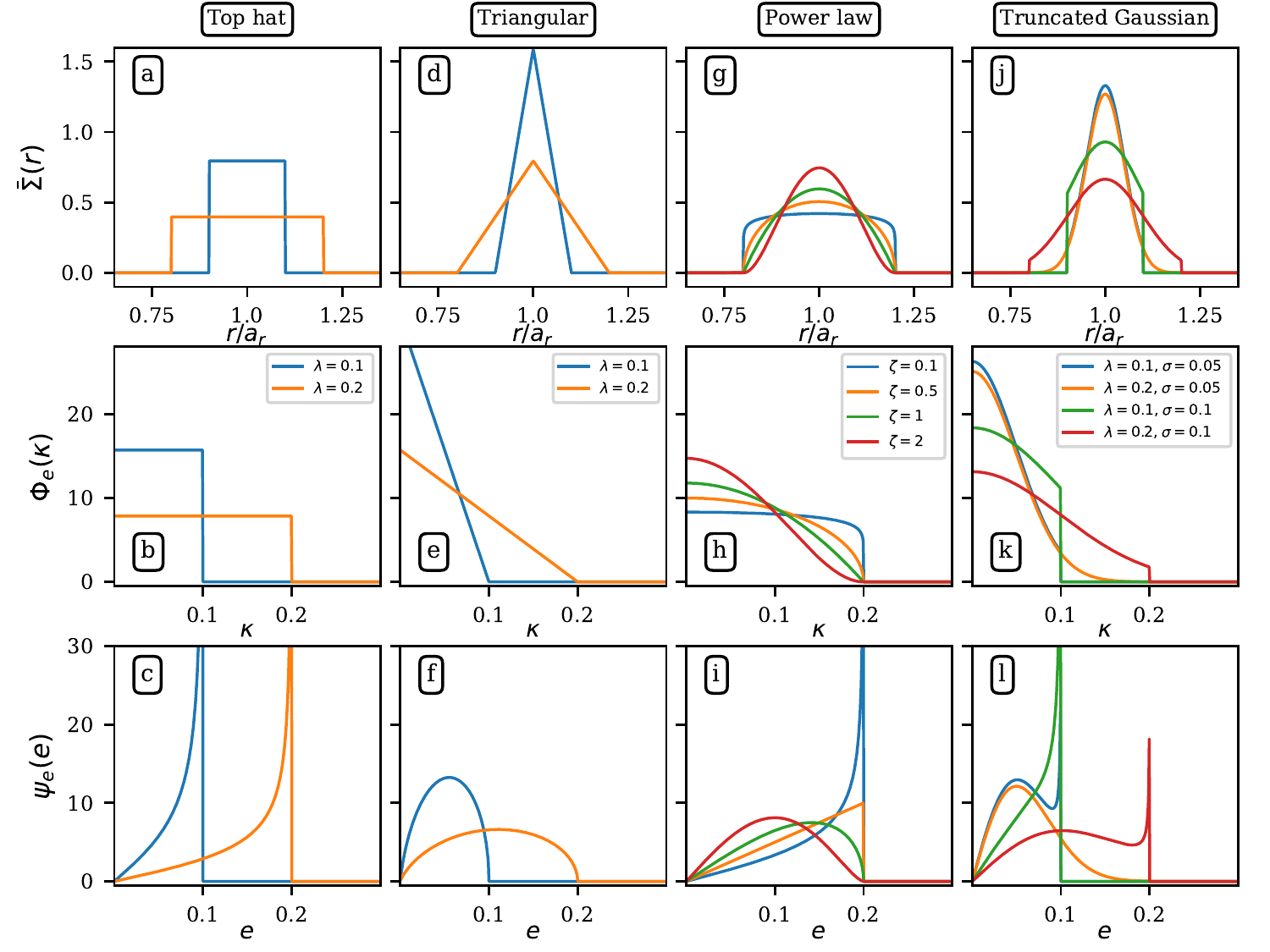}
    \caption{Illustration of the $\psi_e$-reconstruction using $\asd(r)$ profiles of narrow rings. Four columns correspond to the different ASD models: (a)-(c) top-hat, \S\ref{sec:tophat}, (d)-(f) triangular, \S\ref{sec:tri}, (g)-(i) power law, \S\ref{sec:power}, (j)-(l) truncated Gaussian, \S\ref{sec:trunc}. Top, middle and bottom rows show the corresponding $\asd(r)$, $\Phi_e(\kappa)$, and $\psi_e(e)$, correspondingly, for several values of the relevant parameters indicated in the middle row panels. See text for details.}
    \label{fig:ring-profiles}
\end{figure*}

Here we illustrate the process of reconstruction of the explicit form of $\psi_e(e,a)=\psi_e(e,a_r)$ using equation (\ref{eq:Abel}) for several profiles of $\asd(r)$ that might be featured by narrow debris rings. These profiles must necessarily obey the constraint (\ref{eq:sym-ring}), i.e. be symmetric with respect to the ring semi-major axis $a_r$. We will assume the rings to be confined within the radial range of (half-)width $\lambda a_r$, such that 
\ba  
\asd(r)=0~~~~\mbox{for}~~~~|r-a_r|>\lambda a_r.
\ea  
Also, $\asd(r)$ should be normalized such that the total ring mass is $m_r$. 

Given such $\asd(r)$ we will express $\Phi_e(\kappa)$ through the equation (\ref{eq:Psi-ring}), keeping in mind that $\kappa=|r-a_r|/a_r$ in the ring case. After that we will use equation (\ref{eq:Abel}) to obtain $\psi_e(e,a_r)$. Examples of application of this technique follow, providing us with a set of useful  $\psi_e-\Phi_e$ pairs.


\subsection{Ring with a top-hat $\asd(r)$}
\label{sec:tophat}


We start with a top-hat ring of mass $m_r$, for which
\ba
\asd(r)=\frac{m_r}{4\pi a_r^2\lambda}, ~~~~~~|r-a_r|<\lambda a_r,
\label{eq:asd-tophat}
\ea  
(see Figure \ref{fig:ring-profiles}a). Equation (\ref{eq:Psi-ring}) gives for this ASD
\ba
\Phi_e(\kappa)=
\left\{
\begin{array}{ll}
(\pi/2)\lambda^{-1}, & \kappa\le\lambda,\\
0, & \kappa>\lambda.
\end{array}
\right.
\label{eq:Psi-tophat}
\ea   
(see Figure \ref{fig:ring-profiles}b). Since $\Phi_e(\kappa)$ is discontinuous at $\kappa=\lambda$, we explicitly take this into account by writing $\partial\Phi_e/\partial\kappa=-(\pi/2\lambda)\delta(\kappa-\lambda)$. Plugging this into the equation (\ref{eq:Abel}) we deduce
\ba  
\psi_e(e,a_r)=\lambda^{-1}\frac{e}{\sqrt{\lambda^2-e^2}},~~~e<\lambda,
\label{eq:psi-tophat}
\ea  
shown in Figure \ref{fig:ring-profiles}c. Note the divergence of $\psi_e$ near $e=\lambda$.

One can easily verify this form of $\psi_e$ by substituting (\ref{eq:psi-tophat}) into the definition (\ref{eq:Psi_gen}) and recovering $\Phi_e$ given by the equation (\ref{eq:Psi-tophat}). Also, it is easy to check that the distribution function (\ref{eq:psi-tophat}) is normalized to unity, as it should.


\subsection{Ring with a triangular $\asd(r)$}
\label{sec:tri}


A ring with a triangular ASD has (see Figure \ref{fig:ring-profiles}d)
\ba
\asd(r)=\frac{m_r}{2\pi a_r^3\lambda^2}\left(\lambda a_r-|r-a_r|\right), ~~~~~~|r-a_r|<\lambda a_r,
\label{eq:asd-tri}
\ea  
for which (see Figure \ref{fig:ring-profiles}e)
\ba
\Phi_e(\kappa)=
\left\{
\begin{array}{ll}
\pi\lambda^{-2}(\lambda-\kappa), & \kappa\le\lambda,\\
0, & \kappa>\lambda.
\end{array}
\right.
\label{eq:Psi-tri}
\ea  
This $\Phi_e(\kappa)$ is continuous everywhere so the Dirac $\delta$-function no longer appears in $\partial\Phi_e/\partial\kappa$. Equation (\ref{eq:Abel}) gives in this case
\ba  
\psi_e(e,a_r)=\frac{2 e}{\lambda^2}\ln\frac{\lambda+\sqrt{\lambda^2-e^2}}{e},~~~e<\lambda,
\label{eq:psi-tri}
\ea  
see Figure \ref{fig:ring-profiles}f.


\subsection{Ring with a power law $\asd(r)$}
\label{sec:power}


Next we look at an ASD in the "power law" form
\begin{align}
\asd(r) &= \frac{m_r}{2\pi^{3/2} a_r^{2(1+\zeta)}\lambda^{2\zeta+1}}\frac{\Gamma(\zeta+3/2)}{\Gamma(\zeta+1)}
\nonumber\\
&\times  \left[(\lambda a_r)^2 - (r-a_r)^2\right]^\zeta,~~~~~~|r-a_r|<\lambda a_r,
\label{eq:asd-power}
\end{align} 
with $\zeta>0$ being a free parameter; normalization of this expression again ensures that the ring mass is $m_r$. Variation of $\zeta$ allows us to explore a whole class of ASD profiles, continuous at $r=a_r(1\pm \lambda)$, but with rather different characteristics, see Figure \ref{fig:ring-profiles}g. For example, for $\zeta>1$ not only $\asd(r)$ vanishes at $r=a_r(1\pm \lambda)$, but also its derivative, while for $\zeta<1$ the derivative diverges at the ring edges.

This ASD has a corresponding
\ba  
\Phi_e(\kappa)=
\pi^{1/2} \lambda^{-(2\zeta+1)}\frac{\Gamma(\zeta+3/2)}{\Gamma(\zeta+1)}\left(\lambda^2 - \kappa^2\right)^\zeta,
\label{eq:Psi-power}
\ea  
(see Figure \ref{fig:ring-profiles}h) for $\kappa\le\lambda$ (and zero otherwise), which gives, using equation (\ref{eq:Abel}),
\ba  
\psi_e(e,a_r)=(2\zeta+1) \lambda^{-(2\zeta+1)}e\left(\lambda^2 - e^2\right)^{\zeta-1/2},~~~e<\lambda.
\label{eq:psi-power}
\ea  

In particular, as $\zeta\to 0$, ASD (\ref{eq:asd-power}) evolves into a top-hat profile and equations (\ref{eq:asd-power})-(\ref{eq:psi-power}) reduce to the equations (\ref{eq:asd-tophat})-(\ref{eq:psi-tophat}), as expected. For $\zeta=1/2$ (a "semi-circular" ASD profile) one finds
\ba  
\psi_e(e,a_r)=2\lambda^{-2}e,~~~e<\lambda,
\label{eq:psi-semi}
\ea  
while for $\zeta=2$ one obtains
\ba  
\psi_e(e,a_r)=5\lambda^{-5}e\left(\lambda^2 - e^2\right)^{3/2},~~~e<\lambda,
\label{eq:psi-zeta2}
\ea  
see Figure \ref{fig:ring-profiles}i.


\subsection{Ring with a truncated Gaussian $\asd(r)$}
\label{sec:trunc}


We also consider a debris ring with the ASD profile in the form of a truncated Gaussian (see Figure \ref{fig:ring-profiles}j) with a (relative, dimensionless) dispersion $\sigma_\kappa$:
\ba  
\asd(r)=\frac{m_r C}{2\pi^2 a_r^2}\exp\left[-\frac{1}{2}\left(\frac{r-a_r}{\sigma_\kappa a_r}\right)^2\right],
\label{eq:asd-trunc}
\ea  
for $|r-a_r|<\lambda a_r$, where 
\ba  
C=\sqrt{\frac{\pi}{2}}\sigma_\kappa^{-1}\left[\mathrm{erf}\left(\frac{\lambda}{2^{1/2}\sigma_\kappa}\right)\right]^{-1}
\label{eq:C-trunc}
\ea  
guarantees that the ring mass is $m_r$. The corresponding $\Phi_e$ is 
\ba 
\Phi_e(\kappa)=C\exp\left(-\kappa^2/2\sigma_\kappa^2\right),~~~~\kappa\le\lambda,
\label{eq:Psi-trunc}
\ea  
and zero otherwise, thus, discontinuous at $\kappa=\lambda$, see Figure \ref{fig:ring-profiles}k. Similar to the case considered in \S\ref{sec:tophat}, this means that $\partial\Phi_e/\partial\kappa$ has a contribution proportional to $\delta(\kappa-\lambda)$. With this in mind, equation (\ref{eq:Abel}) yields
\begin{align}  
\psi_e(e,a_r)=\sqrt{\frac{2}{\pi}}Ce\Bigg[&
\frac{e^{-e^2/2\sigma_\kappa^2}}{\sigma_\kappa}
\mathrm{erf}\left(\sqrt{\frac{\lambda^2-e^2}{2\sigma_\kappa^2}}\right)
\nonumber\\
& +\sqrt{\frac{2}{\pi}}\frac{e^{-\lambda^2/2\sigma_\kappa^2}}{\sqrt{\lambda^2-e^2}}\Bigg],
\label{eq:psi-trunc}
\end{align} 
which is shown in Figure \ref{fig:ring-profiles}l. In the low-$e$ ($\sigma_\kappa\ll 1$), untruncated ($\sigma_\kappa\ll \lambda$) limit, equations (\ref{eq:psi-trunc}), (\ref{eq:Psi-trunc}) agree with the equations (\ref{eq:Rayleigh}), (\ref{eq:Psi_Ray_app}).

Note the divergent tail of $\psi_e$ as $e\to\lambda$, caused by the sharp truncation of the ring at $r=a_r(1\pm \lambda)$, similar to that in Figure \ref{fig:ring-profiles}c. As indicated by the blue curves corresponding to $\zeta=0.1$ in panels (g)-(i), such divergence of $\psi_e$ is not endemic to discontinuously truncated $\asd(r)$ profiles. It also arises in rings with continuous $\asd(r)$, as long as they decay to zero sufficiently rapidly (i.e. for $\zeta<1/2$ for ASD  (\ref{eq:asd-power})).

It is straightforward to check that all $\psi_e$ derived in this Appendix are normalized to unity and that plugging them back in equation (\ref{eq:Psi_gen}) recovers the corresponding $\Phi_e$. Given the ring setup, with only a single value of $a=a_r$ possible, neither $\psi_e$ nor $\Phi_e$ depend on $a$. But the relations between $\Phi_e$ and $\psi_e$ that we found here are completely general and independent of the semi-major axis distribution $\Sigma_a(a)$ of the debris particles; the ring setup was used just for an illustration (see also \S\ref{sec:psi-edge}). For a general $\Sigma_a(a)$ one would find $\Phi_e$ depending on $a$ at least through the $\kappa(r,a)$ dependence.

Moreover, for general $\Sigma_a(a)$ one can easily introduce some additional $a$-dependence in $\Phi_e$. For example, in all cases considered here $\lambda$ can be made a function of $a$. In equations (\ref{eq:asd-power}) and (\ref{eq:asd-trunc}) one can also make $\zeta$ and $\sigma_\kappa$ to depend on $a$. Introducing such dependencies would not change the {\it mutual} relations between $\psi_e$ and $\Phi_e$ found here, as equations (\ref{eq:Psi_gen}) and (\ref{eq:Abel}) leave them unaffected. 


\section{Reconstruction of $\psi_e$ near a sharp edge}
\label{sec:psi-edge}


We now show an example of $\psi_e$ reconstruction from $\asd(r)$ near a sharp edge of $\Sigma_a(a)$ distribution, see \S\ref{sec:edges-gen}. For that procedure to work, ASD should, in the first place, obey the constraint (\ref{eq:sym-edge}), i.e. $\md\asd/\md\ln r$ should be symmetric with respect to the disc edge at $a_0$. For illustration, let us assume that 
\ba  
\md\asd/\md\ln r=K\left[(\lambda a_0)^2-(r-a_0)^2\right],~~~|r-a_0|\le \lambda a_0,
\label{eq:asd-edge}
\ea  
and zero for $|r-a_0|> \lambda a_0$, which satisfies this constraint; $K<0$ is a constant which will turn out being not a free parameter, see below. This assumption implies that $\asd$ is constant for $|r-a_0|> \lambda a_0$. We set $\asd(r)=0$ for $r>a_0(1+\lambda)$ (i.e. an assumption of an outer edge) and denote $\asd(r)=\Sigma_m$ for $r<a_0(1-\lambda)$, i.e. in the bulk of the disc, unaffected by the edge effects. In general, $\Sigma_m$ is different from $\Sigma_0$ --- the height of plateau of $\Sigma_a$ (equation (\ref{eq:Sig-edge})) --- see Appendix \ref{sec:PL} and equation (\ref{eq:Sigma_m}).

Solving equation (\ref{eq:asd-edge}) with the boundary condition $\asd(a_0(1+\lambda))=0$ we find
\begin{align}
\asd(r)=Ka_0^2\Bigg[&(\lambda^2-1)\ln\frac{a_0(1+\lambda)}{r}-2\left(\frac{r}{a_0}-1-\lambda\right)
\nonumber\\
&+\frac{(r/a_0)^2-(1+\lambda)^2}{2}\Bigg].
\label{eq:asd-edge1}
\end{align} 
Thus, starting with this profile we need to determine the corresponding $\psi_e$ following the recipe outlined in \S\ref{sec:edges-gen} and then using equation (\ref{eq:Abel}).

\begin{figure}
	\includegraphics[width=\linewidth]{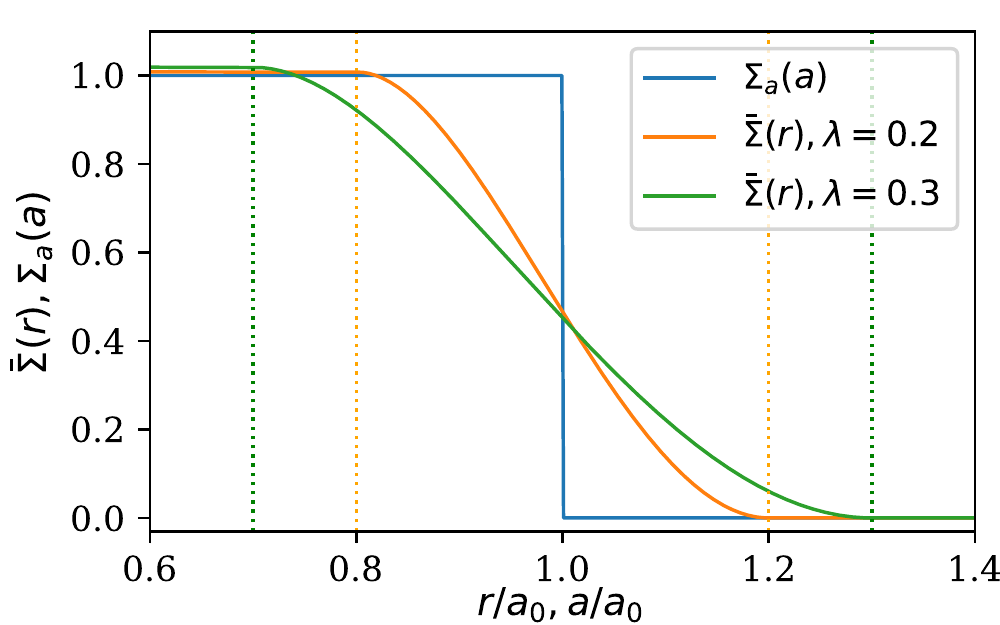}
    \caption{Behavior of $\Sigma_a(a)$ (blue) and $\asd(r)$ given by equation (\ref{eq:asd-edge2}) with $\lambda=0.2$ (orange) and $\lambda=0.3$ (green) near the outer disc edge. Application of the reconstruction described in \S\ref{sec:edges-gen} and Appendix \ref{sec:psi-edge} to this $\asd(r)$ profile results in the eccentricity distribution (\ref{eq:psi-edge}). Dotted vertical lines delineate the interval in which the corresponding curve smoothly transitions from $\asd=0$ to $\asd=\Sigma_m$ given by the equation (\ref{eq:Sigma_m}). }
    \label{fig:edge-rec}
\end{figure}

We start by noting that equation (\ref{eq:asd-edge}) also defines $\Phi_e$ through the equation (\ref{eq:Psi-edge}):
\ba  
\Phi_e(\kappa)=-\frac{\pi K a_0^2}{\Sigma_0}
\left(\lambda^2-\kappa^2\right).
\label{eq:Psi-edge1}
\ea  
This $\Phi_e$ is of the form (\ref{eq:Psi-power}) with $\zeta=1$ (green curve in Figure \ref{fig:ring-profiles}h). Using the results of (\ref{sec:power}), two things follow immediately. First, setting $\zeta=1$ in equation (\ref{eq:psi-power}) we conclude that $\psi_e$ is given by
\ba 
\psi_e(e)=3\lambda^{-3}e\sqrt{\lambda^2-e^2},~~~e<\lambda,
\label{eq:psi-edge}
\ea  
see the green curve in Figure \ref{fig:ring-profiles}i. Second, from the fact that $\psi_e$ in the form (\ref{eq:psi-edge}) is normalized to unity, it immediately follows that $K$ must be such that the equation (\ref{eq:Psi-edge1}) reduces to the equation (\ref{eq:Psi-power}) with $\zeta=1$, i.e.  $K=-(3/4)\lambda^{-3}\Sigma_0/a_0^2$. This implies that our starting ASD profile (\ref{eq:asd-edge1}) must look like 
\begin{align}
\asd(r)=\frac{3\Sigma_0}{4\lambda^3}\Bigg[&(\lambda^2-1)\ln\frac{1+\lambda}{x}-2\left(x-1-\lambda\right)
\nonumber\\
&+\frac{x^2-(1+\lambda)^2}{2}\Bigg],~~~~~~~x=\frac{r}{a_0}.
\label{eq:asd-edge2}
\end{align} 
This immediately gives us $\Sigma_m$ as $\asd(a_0(1-\lambda))$:
\begin{align}
\Sigma_m=\frac{3\Sigma_0}{4\lambda^3}\left[2\lambda-(1-\lambda^2)\ln\frac{1+\lambda}{1-\lambda}\right],
\label{eq:Sigma_m}
\end{align} 
which indeed differs from $\Sigma_0$. Figure \ref{fig:edge-rec} illustrates this profile for two values of $\lambda=0.2, 0.3$, showing in particular that $\asd(r)$ smoothly connects at $r=a_0(1\pm\lambda)$ to the flat segments outside the transition region. 

As $\lambda\to 0$, one finds $\Sigma_m/\Sigma_0=1+(\lambda^2/5)+O(\lambda^4)$, which is reflected in small deviations of $\asd(r)$ from $\Sigma_0$ in the left part of Figure \ref{fig:edge-rec}. In zero-eccentricity limit (i.e. when $\lambda\to 0$, see equation (\ref{eq:psi-edge})) $\asd(r)\to \Sigma_a(r)$, as expected . 


\bsp	
\label{lastpage}

\end{document}